\documentclass[article]{jmlr}% new name PMLR (Proceedings of Machine Learning)
\makeatletter
 \let\Ginclude@graphics\@org@Ginclude@graphics 
\makeatother
\usepackage{pifont}% http://ctan.org/pkg/pifont
\newcommand{\cmark}{\ding{108}}%
\newcommand{\xmark}{\ding{109}}%
\newcommand{\halfmark}{\ding{119}}
 \usepackage{subcaption} 
 \usepackage{booktabs}
 \usepackage{float}
 \usepackage{placeins}
\usepackage{longtable}% for long tables
\usepackage{multirow}
\usepackage{lipsum}

\newcommand\blfootnote[1]{%
  \begingroup
  \renewcommand\thefootnote{}\footnote{#1}%
  \addtocounter{footnote}{-1}%
  \endgroup
}
\makeatletter
\def\set@curr@file#1{\def\@curr@file{#1}} %temp workaround for 2019 latex release
\makeatother
\usepackage[load-configurations=version-1]{siunitx} % newer version

\theorembodyfont{\upshape}
\theoremheaderfont{\scshape}
\theorempostheader{:}
\theoremsep{\newline}

\newcommand{\dvplus}{\textsc{SemanticVessel }}

\title[Fine-grained vessel annotations in CTA]{Scaling up fine-grained intracranial vessel annotations in computed tomography angiography}

\author{\Name{Chu-Hsuan Lin*}
       \Email{clin71@bwh.harvard.edu}\\ 
       \addr Brigham and Women's Hospital,  Harvard Medical School\\
       Boston, MA, USA 
       \AND
       \Name{Alberto Mario Ceballos-Arroyo*}
       \Email{ceballosarroyo.a@northeastern.edu}\\ 
       \addr Khoury College of Computer Sciences,  Northeastern University\\
       Boston, MA, USA
             \AND
       \Name{Jisoo Kim}
       \Email{jkim@bwh.harvard.edu}\\ 
       \addr Brigham and Women's Hospital,    Harvard Medical School\\
       Boston, MA, USA 
       \AND
       \Name{Shrikanth M. Yadav}
       \Email{shrikanth@wustl.edu}\\ 
       \addr Washington University in St. Louis\\
       St. Louis, MI, USA 
       \AND
       \Name{Huaizu Jiang}
       \Email{h.jiang@northeastern.edu}\\ 
       \addr Khoury College of Computer Sciences,  Northeastern University\\
       Boston, MA, USA 
       \AND 
       \Name{Lei Qin}
       \Email{lei\_qin@dfci.harvard.edu}\\ 
       \addr Dana-Farber Cancer Institute,  Harvard Medical School\\
       Boston, MA, USA 
       \AND 
       \Name{Geoffrey S. Young}
       \Email{gsyoung@bwh.harvard.edu}\\ 
       \addr Brigham and Women's Hospital,    Harvard Medical School\\
       Boston, MA, USA 
       } 

\begin{document}

\maketitle

\begin{abstract}

In this work, we present \textsc{SemanticVessel}, a dataset for fine-grained brain vessel segmentation in computed tomography angiography scans. Based on the detailed contrast provided by dynamic 4D-CTA scans, we generate segmentation traces for arteries and veins. We then use intensity-guided region growing to obtain segmentations of the majority of vascular territories in the human brain, which are refined and annotated with 20 unique arterial classes by an expert radiologist. Unlike existing datasets, where minor arteries are discarded as background content, we merge these minor arteries into a generic arterial class. Due to the multiple-phase acquisition of dynamic 4D-CTA, labels for a single phase can be re-used for other phases in the same series, greatly increasing the  size of our dataset with no additional annotation cost. The results show that models trained with the additional generic artery class produce better fine-grained segmentations across the board. We will make our code, annotation GUI, and model weights available to the scientific community.  Code, weights, and data will be made available on \href{https://github.com/alceballosa/robust-vessel-segmentation}{https://github.com/alceballosa/robust-vessel-segmentation}. \blfootnote{* Chu-Hsuan Lin and Alberto Mario Ceballos-Arroyo contributed equally to this work.} 
\end{abstract}

 \thispagestyle{empty}

\section{Introduction}

Intracranial vessel segmentation is a critical task in neuroimaging with broad clinical relevance; the diagnosis and management of cerebrovascular diseases (CVD) such as stroke, intracranial aneurysm, arteriovenous malformation, and cerebral small vessel disease. Accurate distinction of individual vessel segments, including major arteries such as the internal carotid artery (ICA), middle cerebral artery (MCA), anterior cerebral artery (ACA), and posterior cerebral artery (PCA), as well as the venous territories, provides clinicians with quantitative biomarkers for disease assessment, treatment planning, and longitudinal monitoring. Despite the clinical importance of this task, it remains largely performed manually by trained radiologists, a process that is time-consuming, expensive, and subject to inter-rater variability.

Today, intracranial vessels are imaged using various modalities, including computed tomography angiography (CTA), magnetic resonance angiography (MRA), and digital subtraction angiography (DSA). DSA is considered to be the gold standard for some cerebrovascular diseases (CVDs), but its invasive, procedural complication risk, and reliance on X-ray imaging have led to it being used less routinely than the other two~\citep{bash_intracranial_2005}. Compared with MRA, CTA is more cost-effective and widely used in clinical practice due to its accessibility, faster acquisition time, and compatibility with emergency settings. Although MRA offers higher soft-tissue contrast and can provide more detailed visualization of vascular structures without ionizing radiation, its longer scan time and higher cost limit its routine use, particularly in acute scenarios.  In addition, CTA and MRA have been shown~\citep{chen_meta-analysis_2018} to have comparable diagnostic performance for the detection of intracranial aneurysms, with CTA achieving a pooled sensitivity of 0.84 and a specificity of 0.85, and MRA a sensitivity of 0.80 and a specificity of 0.87. Furthermore, ~\cite{phillips_ct_2002} reported that  CTA can outperform MRA for carotid vascular disease detection, with the former showing substantially higher sensitivity than MRA for detecting intracranial stenosis (98\% vs. 70\%) and occlusion (100\% vs. 87\%). As a result, even in centers where MR scanners are  readily available, a single patient will receive scans from both modalities at different points in time based on various criteria. 
 %In contrast, CTA enables rapid and reliable imaging of intracranial vessels, making it a preferred modality for large-scale clinical deployment despite its relatively lower resolution for fine-grained vessel segmentation.

Despite the importance of both modalities in the clinic, most work on intracranial vessel segmentation has focused on MRA data~\citep{hilbert_brave-net_2020,min_automated_2024,zhou_deep_2024}. To address this gap, we propose a unified framework for intracranial vessel semantic segmentation in CTA that jointly delineates and labels both arterial and venous structures at the segment level. Our approach leverages time-resolved CT angiography (4D-CTA), where contrast dynamics naturally distinguish arterial and venous phases. We construct weakly supervised training targets by combining binary vessel segmentation with semantic labels through proximity-based propagation. To the best of our knowledge, this is the first time end-to-end semantic segmentation of intracranial cerebrovascular vessels is achieved, simultaneously identifying and labeling both arteries and veins within a single model.

\subsection*{Generalizable Insights about Machine Learning in the Context of Healthcare}
%This section is \emph{required}, must keep the above title, and should
%be the final part of your introduction.  In about one paragraph, or
%2-4 bullet points, explain what we should \emph{learn} from reading
%this paper that might be relevant to other machine learning in health
%endeavors.

%For example, a work that simply applies a bunch of existing algorithms
%to a new domain may be useful clinically but doesn't increase our
%understanding of the machine learning and healthcare; if that study
%also investigates \emph{why} different approaches have different
%performance, that might get us excited!  A more theoretical machine
%learning work may be in how it enables a new kind of clinical study.
%\emph{Reviewers and readers will look to evaluate (a) the significance
%  of your claimed insights and (b) evidence you provide later in the
%  work of you achieving that contribution}
Semantic segmentation is widely used in medical imaging, with applications such as organ and brain segmentation ~\citep{meijs_robust_2017, wasserthal_totalsegmentator_2023,ma_segment_2024} playing an important role in clinical practice. In intracranial vessel segmentation, automated semantic segmentation can assist in the diagnosis of cerebrovascular diseases and reduce radiologist workload. Automatically-generated brain vessel segmentations can also be very useful to contrastively pre-train vision-language models so that they better recognize the brain vasculature. However, many public datasets do not adequately capture very distal vessels due to limited image contrast and the small size of these structures. As a result, during model training, distal vessels are often lumped together with background parenchyma. Recent work ~\citep{saluja_backsplit_2025} has shown that combining many structures under a generic background label can hamper a model's capabilities. Inspired by this finding, we annotate as many veins as possible and we label the majority of important brain arteries while retaining an “undefined” class to account for arterial regions that would be too costly to annotate in detail. Our experiments indicate that Dynamic 4D-CTA data can significantly increase the amount of brain CTA that can be annotated with fine-grained labels by a single neuro-radiologist, shifting the majority of the burden to the relatively simple task of assigning labels to segmented vessel segments. We hope that our work will inspire the development of new and more comprehensive datasets for intracranial vessel segmentation, as well as novel contrastive pre-training strategies for cerebrovascular data.
%Make sure you also put your work in the context of related
%work.  Who else has worked on this problem, and how did they approach
%it?  What makes your direction interesting or distinct?

\section{Related Work}

\subsection{Annotating and Segmenting Brain Vessels in CTA Scans}

Most prior work approached intracranial vessel segmentation as a binary task, distinguishing vessels from background tissue ~\citep{meijs_robust_2017, fu_rapid_2020, patel_evaluating_2023, van_voorst_deep_2026}, resulting in a unified vascular mask without anatomical differentiation. Some studies aiming to incorporate anatomical specificity have followed two main approaches:

The first involves direct multiclass voxel-wise segmentation; the TopCoW challenge~\citep{yang2025topcow} established the first public benchmark for 13-class Circle of Willis (CoW) segmentation across both MRA and CTA modalities, with topology-aware approaches subsequently improving structural consistency within this region~\citep{zhang2024topcow24,hamadache2026topology}. However, these efforts are confined to the Circle of Willis, limiting their applicability to other parts of the brain. The recently announced TopBrain challenge~\citep{yang_topbrain_2026} extends the annotation scope to over 40 vessel labels spanning both arteries and veins across CTA and MRA, but only 25 scans per modality are available.

The second direction treats anatomical identification as a post-hoc labeling problem applied on top of a binary segmentation. Some of these approaches rely on graphs, assigning labels to bifurcation nodes or centerline segments via graph matching~\citep{rist2023bifurcation, thamm2022labeling} or deep learning on preprocessed vessel masks~\citep{chen2024cta}. Other approaches have attempted to assign labels to segments by registering arterial atlases to the target scans, but their use remains limited to MRA data~\citep{falcetta2026automatedframeworklargescalegraphbased}. Overall, while covering a wider anatomical scope beyond the Circle of Willis, these methods are two-stage pipelines that depend on the quality of the upstream binary segmentation.

\subsection{Gaps in existing annotation approaches}

Existing large-scale annotation efforts for brain CTA usually produce annotations made from scratch, meaning experts have to review and annotate hundreds of individual slices. 3D augmented reality tools have been developed and used in some cases~\citep{yang2025topcow, yang_topbrain_2026}. However, due to the high workload of neuro-radiologists, the amount of data that can be annotated is limited, and in the case of TopBrain, the performance of published methods is constrained by size of the dataset. The VesselVerse~\citep{FalDan_VesselVerse_MICCAI2025} initiative tackles the annotation availability issue by proposing an integrative framework that combines annotations from multiple experts and models on various CTA and MRA datasets, but it remains limited to binary vessel-background segmentations.

Due to the superior contrast and background-vessel separability of MRA, comprehensive annotation of such datasets is far easier. For this modality, tools such as iCafe~\citep{chen_quantification_2019}, provide a relatively fast annotation framework where the neuro-radiologist only has to assign labels to individual segments, but they are not directly usable with CTA scans and require significant contrast enhancement for this modality. Recently, it has been proposed to subtract baseline images from arterial- and venous-phase images in Dynamic 4D-CTA datasets to simulate the contrast provided by MRA, enabling the use of iCafe for CTA~\citep{ceballos2026dynavessels}. However, iCafe introduces a complicated, non-open-source dependency, limiting the reproducibility of such an approach. %Moreover, in~\citep{ceballos2026dynavessels}, fine-grained segmentation capabilities are not properly benchmarked, with evaluations instead centering on how well individual segments are contained within a two-way arterial/venous segmentation mask.

\section{Methods}

%Tell us your techniques!  If your paper is develops a novel machine
%learning method or extension, then be sure to give the technical
%details---as you would for a machine learning publication---here and,
%as needed, in appendices.  If your paper is developing new methods
%and/or theory, this section might be several pages.

%If you are combining existing methods, feel free to cite other
%packages and papers and tell us how you put them together; that said,
%the work should stand alone for someone in that general machine
%learning area.  

%\emph{Lack of technical details, such that the soundness of the
%  methods can be verified, is a major reason that otherwise
%  strong-looking papers are scored low/rejected.}
\begin{figure}
\centering
\includegraphics[width=0.8\linewidth]{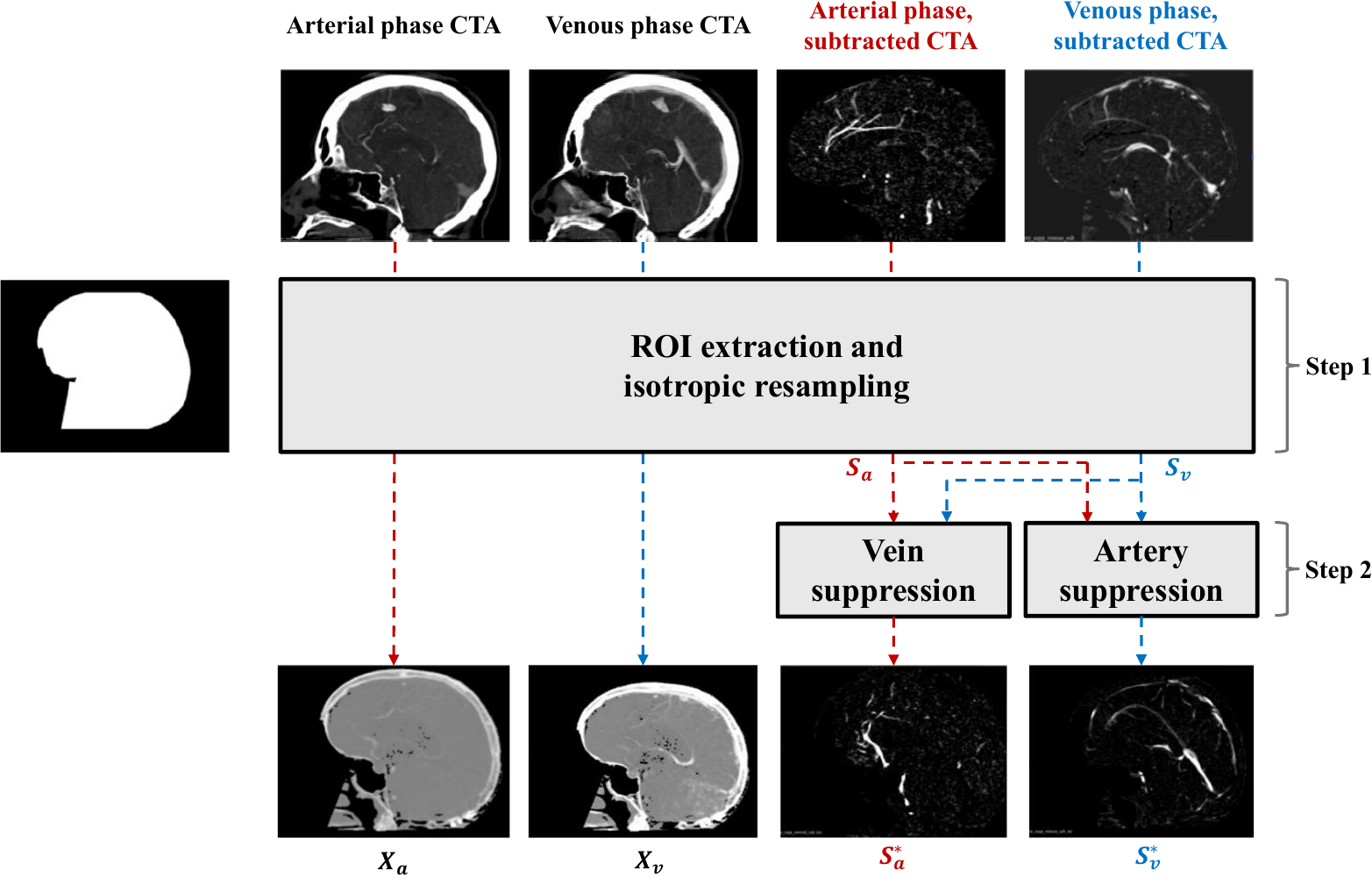}
\caption{Pre-processing steps for a dynamic CTA study. For both phases, a head ROI is used on the CTA and the subtracted CTA to suppress the background. Next, the subtracted images are processed to create the vessel-separated images. }
\label{fig:dynacta-preprocessing}
\end{figure}

\subsection{Dynamic CTA preprocessing}
To obtain better contrast arterial and venous structures, a two-step pre-processing pipeline was applied to the dynamic CTA images, as illustrated in Fig.~\ref{fig:dynacta-preprocessing}. The pipeline takes as input the arterial ($X_a$) and venous ($X_v$) CTA images with their scanner-generated subtracted counterparts ($S_a$, $S_v$). In the first step, each phase is registered to a common spatial reference frame to correct for inter-acquisition motion; taking the arterial phase as the reference, the venous CTA image $X_v$ is affine registered to $X_a$, and the same transformation is applied to $S_v$ to obtain $S_{v \rightarrow a}$, the subtracted venous image is warped into the arterial space. A skull-and-neck binary mask, derived from a CT atlas \citep{talou_adaptive_2021} via affine registration, is additionally applied to exclude extracranial voxels from all images. The same procedure is repeated for the venous phase to obtain $S_{a \rightarrow v}$.

In the second step, to enhance vascular contrast and eliminate erroneous highlights from the opposing phase, separate vein and artery suppression are performed. Specifically, vein suppression is applied to the arterial phase by retaining voxels where $S_a - S_{v \rightarrow a} > 0$ and zeroing out those where $S_a - S_{v \rightarrow a} \leq 0$, yielding the bone- and vein-suppressed volume $S^*_a$. Artery suppression is applied to the venous phase, retaining voxels where $S_v - S_{a \rightarrow v} > 0$ and zeroing out those where $S_v - S_{a \rightarrow v} \leq 0$, producing the bone- and artery-suppressed volume $S^*_v$.

\subsection{Ground Truth Generation}

%Manual voxel-level of intracranial vessels is impractical due to the time and expertise required. We therefore developed an automated pipeline to generate weak ground truth using vessel tracing, region growing, and centerline-based label propagation.
Building on the pre-processing pipeline described above, the bone- and vein-suppressed arterial volume $S^{*}_{a}$ and the bone- and artery-suppressed venous volume $S^{*}_{v}$ were used as the arterial and venous images, respectively.

{To generate seed candidates for region growing, each volume was first normalized using Nyul histogram normalization to standardize intensity distributions across cases. Renyi entropy-based auto-thresholding was then applied to generate an initial binary vessel mask. The multi-scale Hessian vesselness filter was subsequently applied to enhance tubular structures, with sigma range $[0.5, 2.0]$ mm with 4 steps, $\alpha=0.5$, $\beta=1.0$, $\gamma=5.0$. The vesselness response was normalized to $[0, 1]$ and thresholded at $0.2$ to ensure vessel connectivity, yielding the binary vessel masks $T^{*}_{a}$ and $T^{*}_{v}$ with small connected components removed.}

To recover missed thin vessels, intensity-guided region growing was applied to $S^{*}_{a}$ and $S^{*}_{v}$. Seed voxels were selected from $T^{*}_{a}$ and $T^{*}_{v}$ and ranked by their local neighborhood maximum intensity. For each seed, a tolerance was defined as the difference between the local neighborhood maximum and the seed voxel intensity, and was constrained to lie within $\left[\frac{1}{5}, \frac{1}{2}\right]$ of the neighborhood maximum to prevent over- or under-segmentation. Seeds failing this criterion were skipped and the next candidate was evaluated. {To ensure spatial diversity, candidate seeds located too close to previously accepted seeds were excluded from further consideration. Seed spacing was set to $5$ voxels with neighborhood size of $7$ voxels for arteries and $10$ voxels for veins, reflecting the larger cross-sectional diameter of venous structures compared to arteries.} The region growing process continued iteratively until the segmented region remained below $1.5\%$ of the total volume for venous structures and $1\%$ for arterial structures. In cases where the default region-growing parameters produced suboptimal results, assessed by visual inspection on the source CTA image, parameters were adjusted, and the pipeline was rerun.

\begin{figure}
\includegraphics[width=1\linewidth]{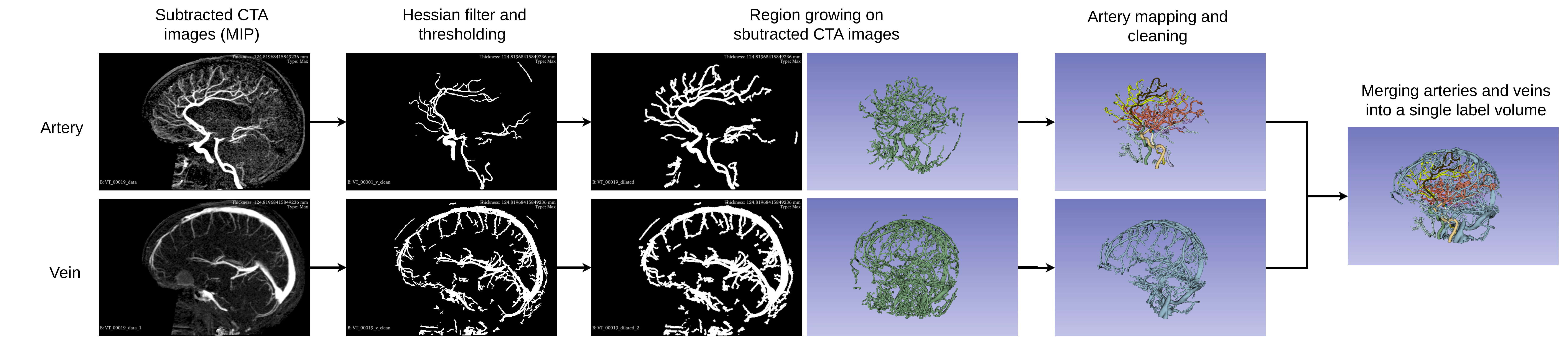}
\caption{The proposed methodology for semi-automated brain vessel annotation.}
\label{fig:gt_flowchart}
\end{figure}

The artery and vein masks were then merged, and a brain mask~\citep{wasserthal_totalsegmentator_2023} was applied to remove non-brain arterial components and retain only the largest venous component, with arteries prioritized in overlapping regions. The full process is illustrated in Figure~\ref{fig:gt_flowchart}.

\begin{figure}
\includegraphics[width=1\linewidth]{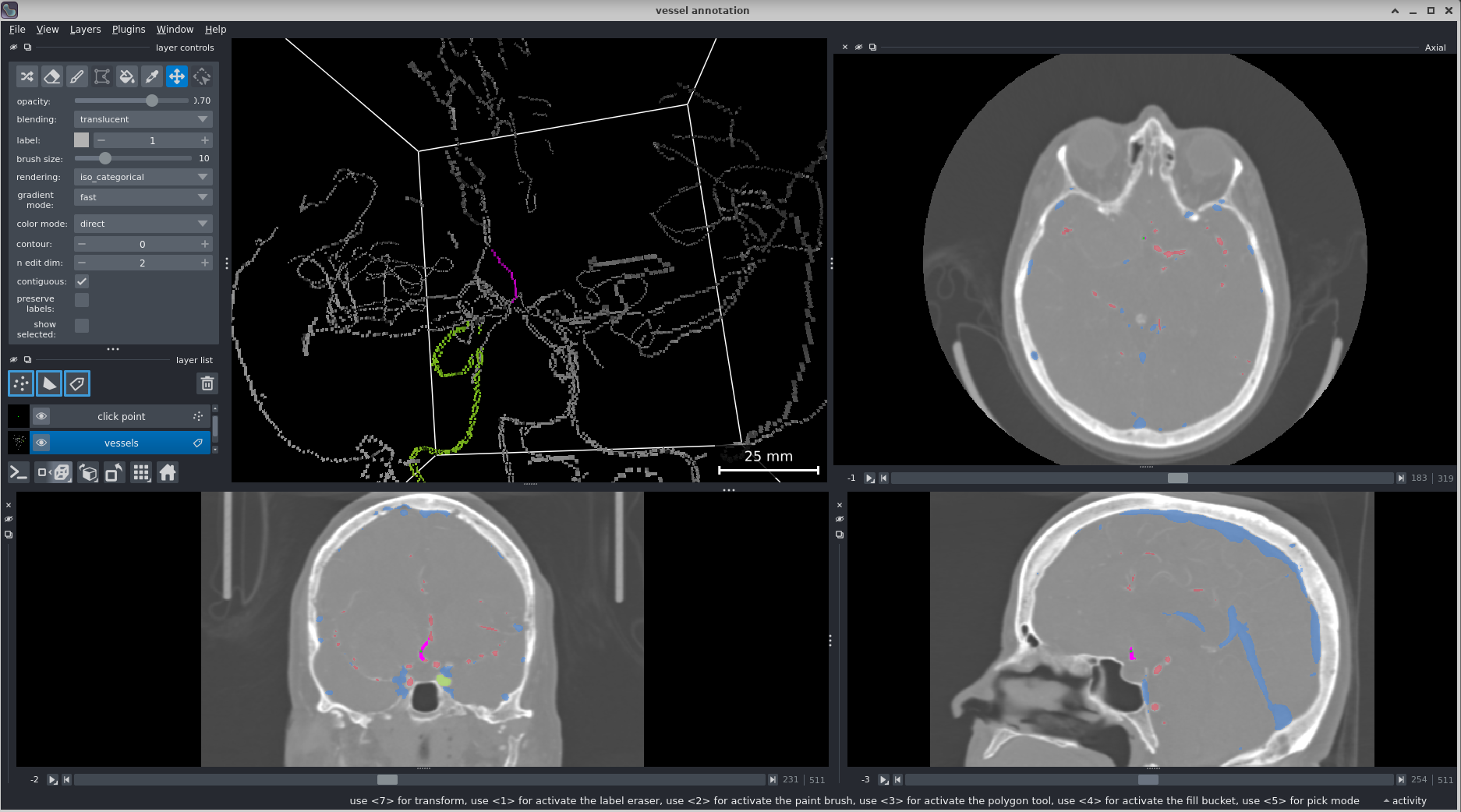}
\caption{Annotation graphical user interface (GUI) for vessel labeling. 
Arteries and veins are displayed in \textcolor{red}{red} and 
\textcolor{blue}{blue} by default, respectively. Once labeled, each vessel 
is assigned a unique color corresponding to its vessel name (e.g., 
\textcolor{green}{green}). Clicking on a vessel highlights it simultaneously 
in both the 3D rendering and 2D slice views in 
\textcolor{magenta}{bright magenta}, enabling consistent cross-view 
identification during the annotation process.}
\label{fig:annotation-gui}
\end{figure}

{For semantic labeling, a skeletonization algorithm was applied to the binary arterial masks to extract vessel centerlines. An experienced neuro-radiologist with 5 years of subspecialty experience then manually assigned 20 arterial class labels to the extracted centerlines using our in-house developed interactive GUI (Figure~\ref{fig:annotation-gui}), which supports both 3D and 2D centerline-based label assignment as well as brush-based tools for adding or erasing vessel segmentations where necessary. The radiologist additionally reviewed and confirmed the quality of the underlying segmentation prior to finalizing the annotations.} The classes are ICA (R/L), M1 (R/L), M2+ (R/L), A1 (R/L), A2+ (R/L), AComm, VA (R/L), BA, P1 (R/L), P2+ (R/L), and PComm (R/L). Labels were assigned to artery voxels based on the nearest labeled voxel within a 2.5~mm radius. Voxels beyond this threshold were assigned to an ``other artery'' class, while all venous voxels were grouped into a single ``vein'' class, resulting in a total of 22 classes. Finally, all phases of the dynamic CTA were rigidly registered to the arterial phase using ANTs~\citep{tustison_antsx_2021}, and the same label map was propagated across phases. In total, we produced semi-automatic fine-grained annotations for 41 patients with a total 360 CTA volumes. In addition, the radiologist manually annotated the vascular tree for 5 held-out patients with a total of 50 scans using 3D Slicer~\citep{Kikinis2014}, based off their peak arterial and venous phase scans. %A comparison with existing CTA brain vessel segmentation datasets is provided in Table \ref{tab:splits}.

%For semantic labeling, 20 arterial classes were defined based vessel traces manually defined on top of the artery mask by a neuro-radiologist with 5 years of subspecialty experience. The classes are ICA (R/L), M1 (R/L), M2+ (R/L), A1 (R/L), A2+ (R/L), AComm, VA (R/L), BA, P1 (R/L), P2+ (R/L), and PComm (R/L). Labels were assigned to artery voxels based on the nearest labeled voxel within a 2.5~mm radius. Voxels beyond this threshold were assigned to an ``other artery'' class, while all venous voxels were grouped into a single ``vein'' class, resulting in a total of 22 classes. Finally, all phases of the dynamic CTA were rigidly registered to the arterial phase using ANTs~\cite{tustison_antsx_2021}, and the same label map was propagated across phases. In total, we produced fine-grained annotations for 41 patients with a total 360 CTA volumes. A comparison with existing CTA segmentation datasets is provided in Table \ref{tab:splits}.

\subsection{Training}
We trained a 3D segmentation network using the nnU-Net framework~\citep{isensee_nnu-net_2021} 
with the Residual Encoder Large (ResEncL) preset. 
The model was trained on all available cases for 1,000 epochs using stochastic gradient 
descent (SGD) with an initial learning rate of $0.01$, momentum of $0.99$, 
and weight decay of $3 \times 10^{-5}$. The network receives two input channels: the CTA image and a subject-specific left--right coordinate volume, in which each voxel encodes its normalized distance relative to the brain midline, ranging from $-1$ (left) to $+1$ (right). This coordinate volume provides the model with explicit spatial context regarding hemispheric laterality, enabling it to distinguish anatomically homologous structures on opposite sides of the brain, which is particularly important for paired vessels such as the left and right middle cerebral arteries. The base training loss combines the Dice and cross-entropy losses. To improve the connectivity and topological completeness of predicted vessel segments, we also trained a second model with skeleton recall loss~\citep{kirchhoff_skeleton_2024}, which penalizes missed centerline voxels more heavily and thereby better preserves vessel connectivity. 

\subsection{Post Processing}

For all model outputs, we apply a per-class threshold to remove residual segmentations with under 100 voxels. This helps remove noisy predictions for subjects with anatomical variations or occluded arteries.

\section{Cohort}
\textbf{Internal dynamic CTA Dataset} This retrospective study was approved by the Institutional Review Board with a waiver of informed consent (MGB HRC IRB Protocol \#2022P000792). The 4D-CTA scans were retrieved from our clinical research database and acquired on a 320-detector-row CT scanner (Canon-Toshiba Aquilion One). Acquisition parameters were as follows: 80~kV tube voltage, 150~mA tube current, 0.75~s rotation time, 320 slices at 0.5~mm thickness, and a $512 \times 512$ matrix. For each patient, scanning commenced approximately 7 seconds following the injection of 75--100~mL of iodinated contrast agent (GE Omnipaque 350) administered at a flow rate of 4--5~mL/s, and continued intermittently across the baseline, arterial, and venous phases.

\begin{figure}
    \centering
    \includegraphics[width=1\linewidth]{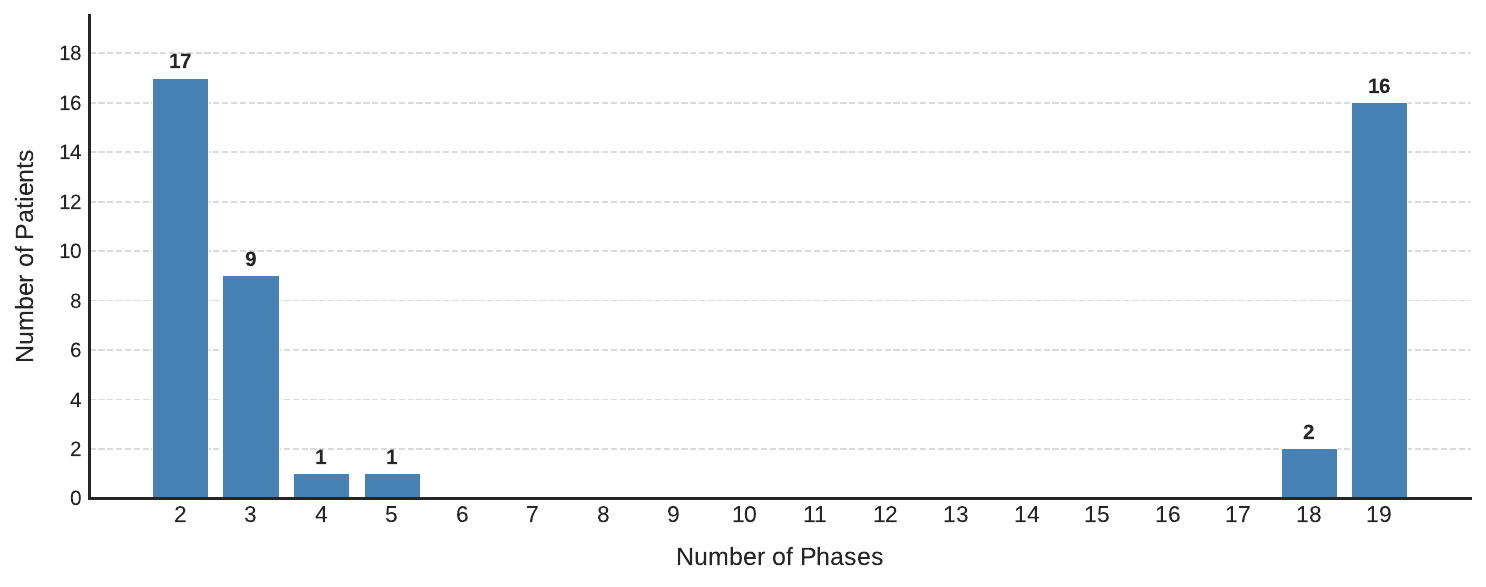}
    \caption{Number of dynamic CT phases per patient in the dataset, including test data (n=46).}
    \label{fig:phase_distribution}
\end{figure}

A total of 1,910 patients scanned between 2016 and 2022 were initially retrieved. Among these, 68 patients had dynamic CTA studies containing at least one peak arterial and one peak venous phase at acceptable resolution, including both original post-contrast images and corresponding subtracted images. Many cases additionally contained intermediate temporal phases, resulting in variable phase counts across the cohort. The distribution of phases is illustrated in Fig~\ref{fig:phase_distribution}. From this cohort, 41 patients were randomly selected for semi-automatic annotation and model training. A further 5 patients scanned between 2023 and 2025 were chosen for our held-out test subset.

\begin{figure}   
    \includegraphics[width=1\linewidth]{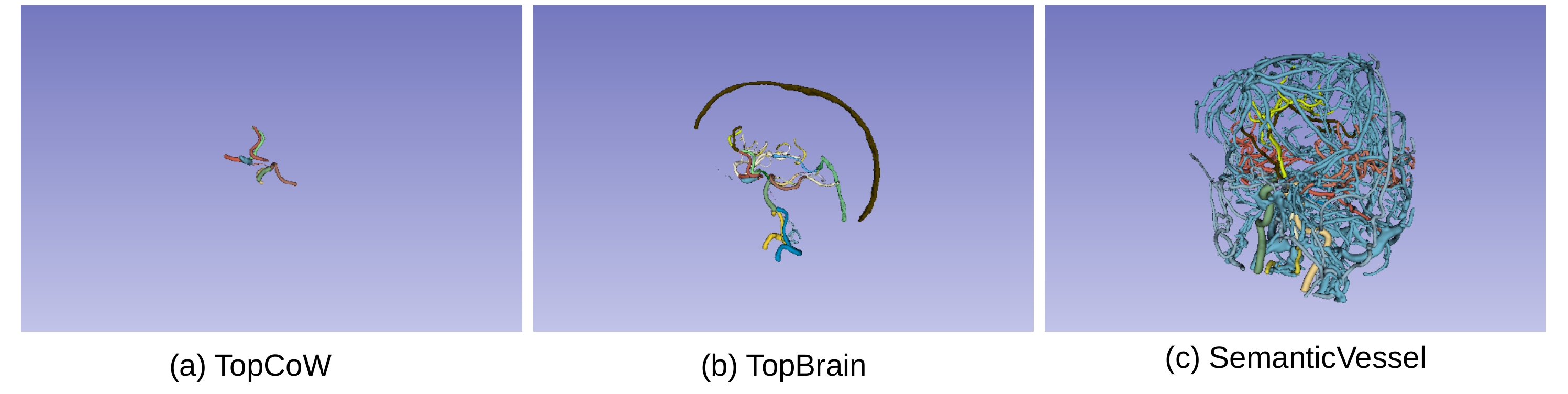}
    \caption{Example of our proposed brain vessel segmentations, compared with the public datasets TopCoW \citep{yang2025topcow} and TopBrain \citep{yang_topbrain_2026}}
    \label{fig:example}
\end{figure}

\textbf{TopCoW and TopBrain Datasets}
The TopCoW ~\citep{yang2025topcow} and TopBrain ~\citep{yang_topbrain_2026} datasets are publicly available neurovascular imaging datasets that provide paired MRA and CTA scans with annotated intracranial vessels. In this study, the 125 CTA images from TopCoW and the 25 CTA images from TopBrain are used as out-of-domain evaluation datasets\footnote{The VesselVerse dataset contains no fine-grained vessel annotations, and thus is omitted from our evaluations.}. Table~\ref{tab:splits} and Figure~\ref{fig:example}
show the differences in coverage and size between  the three datasets.

\begin{table}[]
\centering
\tiny
\caption{Description of existing datasets for brain vessel segmentation in CTA. Coverage for contrast phases (Ctrs.), proximal arteries (Pr.), first (1O) and second (2O) order arteries, distal arteries (Di.), arterial branches (Br.) and veins (V.) is indicated as \cmark: complete or near-complete, \halfmark: partial, or \xmark: absent. We only include training partitions when describing dataset size.
\label{tab:splits}} 
\begin{tabular}{lccccccccc}
\hline
\multirow{2}{*}{\textbf{Dataset}} & \multirow{2}{*}{\textbf{\# scans}} & \multirow{2}{*}{\textbf{Labels}} & \multicolumn{7}{c}{\textbf{Coverage}}    \\ \cline{4-10} 
                                  &                                                                             &                                  & \multicolumn{1}{c}{\textbf{Ctrs.}} & \multicolumn{1}{c}{\textbf{Pr.}} & \multicolumn{1}{c}{\textbf{1O}} & \multicolumn{1}{c}{\textbf{2O}} & \multicolumn{1}{c}{\textbf{Di.}} & \multicolumn{1}{c}{\textbf{Br.}} & \textbf{V.}\\ \hline
TopCoW~\citep{yang2025topcow}                             & 125                                                                              & 13 arteries                      & \multicolumn{1}{c}{\xmark}                    & \multicolumn{1}{c}{\cmark}         & \multicolumn{1}{c}{\cmark}               & \multicolumn{1}{c}{\xmark}                 & \multicolumn{1}{c}{\xmark}            & \multicolumn{1}{c}{\xmark}              & \xmark           \\ \hline
VesselVerse~\citep{FalDan_VesselVerse_MICCAI2025}                       & 125                                                                               & Vessel-only                      & \multicolumn{1}{c}{\xmark}                    & \multicolumn{1}{c}{\cmark}         & \multicolumn{1}{c}{\cmark}               & \multicolumn{1}{c}{\cmark}               & \multicolumn{1}{c}{\xmark}            & \multicolumn{1}{c}{\xmark}              & \halfmark \\ \hline
TopBrain~\citep{yang_topbrain_2026}                           &  25                                     &  6 veins / 34 arteries          & \multicolumn{1}{c}{\xmark}                    & \multicolumn{1}{c}{\cmark}         & \multicolumn{1}{c}{\cmark}               & \multicolumn{1}{c}{\cmark}               & \multicolumn{1}{c}{\cmark}          & \multicolumn{1}{c}{\cmark}            & \cmark         \\ \hline
DynaVessel~\citep{ceballos2026dynavessels}                           &  110                                     &  Vein / Artery          & \multicolumn{1}{c}{\cmark}                    & \multicolumn{1}{c}{\cmark}         & \multicolumn{1}{c}{\cmark}               & \multicolumn{1}{c}{\cmark}               & \multicolumn{1}{c}{\cmark}          & \multicolumn{1}{c}{\cmark}            & \cmark         \\ \hline
\dvplus                     & 360                                                                           & Vein / 22 arteries            & \multicolumn{1}{c}{\cmark}                  & \multicolumn{1}{c}{\cmark}         & \multicolumn{1}{c}{\cmark}               & \multicolumn{1}{c}{\cmark}               & \multicolumn{1}{c}{\cmark}            & \multicolumn{1}{c}{\cmark}              & \cmark         \\ \hline
\end{tabular}
\end{table}

\section{Results} 

%\emph{Depending on the claim you make in the paper, different
%  components may be important for this section.}

\subsection{Evaluation Approach/Study Design} 
%Before jumping into the results: what exactly are you evaluating?
%Tell us (or remind us) about your study design and evaluation
%criteria.
To investigate the impact of label completeness on segmentation performance, we evaluate \textbf{four} label configurations: (1) All labels: training with all labeled structures, including arteries and veins; (2) All arteries: excluding the \textit{vein} class; (3) Named arteries: excluding both the \textit{other artery} and \textit{vein} classes; and \textbf{(4) Ignore labels: treating both the \textit{other artery} and \textit{vein} voxels as ignored labels that are not taken into consideration, such that these regions do not contribute to the loss during training.} To isolate the effect of each variable, we adopt a sequential evaluation strategy. First, we vary the label configuration while keeping the loss function fixed to the default nnU-Net compound loss (Dice and Cross-Entropy), training one model per configuration for a total of three models. Based on these results, we identify the best-performing label configurations and investigate the effect of the loss function by additionally training two models with skeleton recall loss~\citep{kirchhoff_skeleton_2024}.

{In addition to validating the generalizability of our annotation pipeline, we conduct two complementary experiments: (1) Arterial phase only: training an additional model using only the arterial phase images and comparing its performance against the multiphase model; and (2) Cross-dataset training: training on a subset of the TopBrain CTA scans and evaluating the TopBrain, \textsc{SemanticVessel} multiphase, and \textsc{SemanticVessel} arterial phase-only models on 5 held-out TopBrain cases and 5-held-out \textsc{SemanticVessel} cases.}

\subsection{Evaluation Metrics}

As established before, we evaluate our trained models on TopCoW~\citep{yang2025topcow} and TopBrain~\citep{yang_topbrain_2026}. Since the label definitions in these datasets differ from those used in our training data, we establish a unified label scheme by merging semantically equivalent classes across datasets prior to evaluation. The full details of the label mapping are provided in the Appendix~\ref{tab:topbrain_topcow_semanticvessel}.

We assess segmentation performance using three complementary metrics designed to capture volumetric overlap, boundary accuracy, and topological  completeness, respectively. Let $A_i$ denote the ground truth volume, $P$ the predicted volume,
$A_{s,i}$ and $P_s$ the corresponding surfaces, and $A_{c,i}$ the 
centerline volume of $A_i$, we define the three metrics as follows:

\noindent\textbf{(1) Modified Dice Coefficient (mDC)} measures the 
sensitivity of the predicted segmentation with respect to the ground 
truth, and is defined as:
\begin{equation}
    \text{mDC}(A_i, P) = \frac{|A_i \cap P|}{|A_i|}
\end{equation}
Unlike the standard Dice coefficient, mDC is asymmetric and penalizes 
false negatives rather than false positives, making it more suitable 
for evaluating recall of small and distal vessels.

\noindent\textbf{(2) Average Directed Hausdorff Distance (adHD)} 
measures the mean surface distance from the ground truth boundary 
to the nearest predicted boundary point:
\begin{equation}
    \text{adHD}(A_{s,i}, P_s) = \frac{1}{|A_{s,i}|} \sum_{a \in A_{s,i}} 
    \min_{p \in P_s} d(a, p)
\end{equation}
where $d(a, p)$ is the Euclidean distance between surface points.
We adopt the directed variant~\cite{} rather than the symmetric 
Hausdorff distance, as the predicted segmentation boundary can be 
substantially larger than the ground truth in vessel segmentation. Compared with mDC, the adHD metric penalizes segmentations that are too thick compared with the ground truth.

\noindent\textbf{(3) Topology Sensitivity (tSens)} quantifies the 
fraction of the ground truth centerline captured by the predicted 
segmentation, following clDice~\citep{shit_cldice_2021}:
\begin{equation}
    \text{tSens}(A_{c,i}, P) = \frac{|A_{c,i} \cap P|}{|A_{c,i}|}
\end{equation}
A high tSens indicates that the predicted segmentation preserves 
the topological connectivity of the vessel tree, which is critical 
for downstream clinical analysis.

\subsection{Results on Fine-Grained Vessel Segmentation} 

\begin{figure}
    \centering
    \includegraphics[width=1\linewidth]{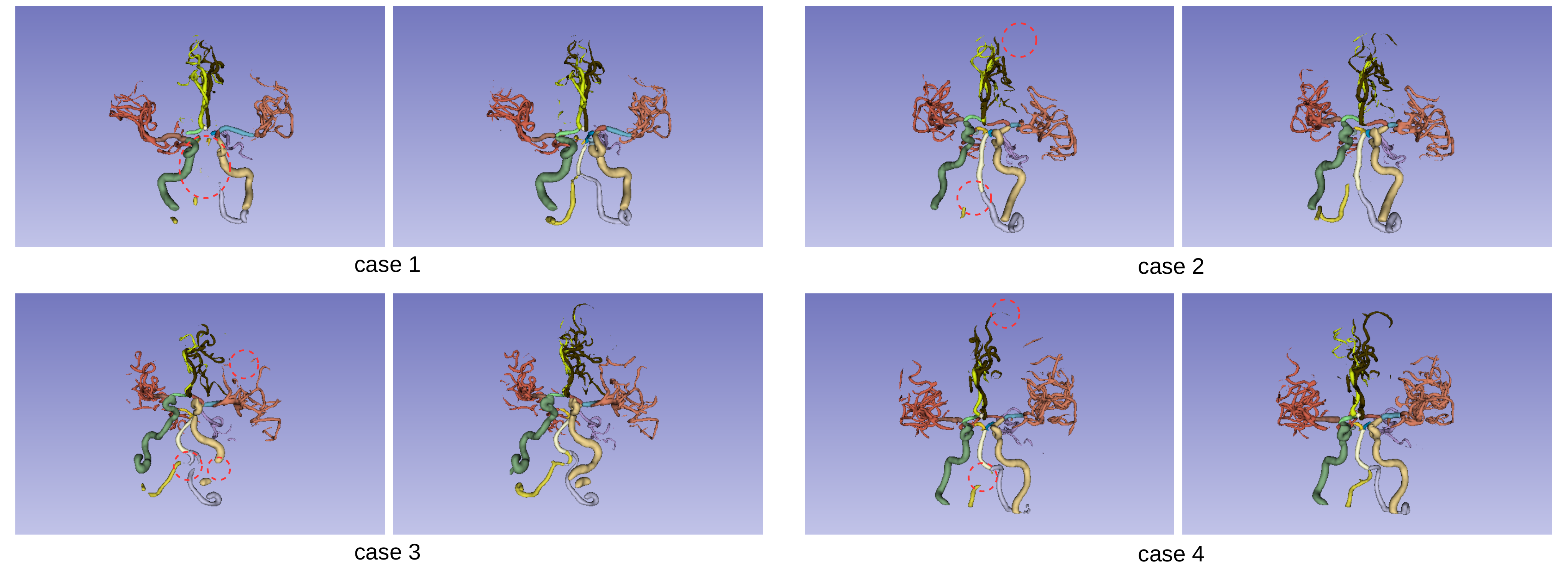}
    \caption{Comparison between the named arteries (left) and all labels (right) training configurations. Red dashed circles indicate regions of missing predictions in the model that was only trained on the named arteries.}
    \label{fig:compare_all_artery}
\end{figure}

All models were trained and tested on a workstation with an NVIDIA 4090 GPU (24 GB VRAM) and 8 CPU cores. Table~\ref{tab:aggregate_combined} reports the aggregated results across three experimental settings, evaluating model performance trained with Dice Loss and Cross-Entropy (CE) Loss and Skeleton Recall Loss on the TopBrain dataset. For Dice Loss and Cross-Entropy (CE) Loss, across all three metrics, the \textit{Named arteries} configuration establishes a baseline, while progressively incorporating additional structures --- extra arteries and veins --- leads to consistent improvements across all metrics. Notably, the \textit{All arteries} configuration (named arteries + other artery) accounts for the majority of the performance gain, suggesting that the extra artery class provides the most significant supervisory benefit. \textbf{ The \textit{All arteries} configuration achieves the highest artery-only mDC of $0.8847 \pm 0.0407$. When further incorporating the vein class in the \textit{All labels} configuration, the artery mDC decreases slightly ($0.8734 \pm 0.0388$),  but we achieve the highest tSens  ($0.9387 \pm 0.0257$) and lowest adHD ($0.2399 \pm 0.1289$) among all configurations, indicating that adding veins provides complementary benefit without sacrificing artery segmentation performance.} A detailed per-label breakdown result for the Skeleton Recall Loss can be found in Table~\ref{tab:per_label_dice}. Figure~\ref{fig:compare_all_artery} depicts a comparison between the \textit{Named arteries} and \textit{All labels} training configurations. While differences in distal vessels are subtle, the \textit{Named arteries} model notably fails to segment major arteries such as the VA and ICA.

When using Skeleton Recall Loss, we see a slight improvement across the board when compared to the model trained using Dice and CE Loss, with the relative ranking of vessel classes and label configurations remaining stable across both training objectives. The vein class continues to be segmented very well, while large and well-opacified vessels such as the M2+ branches maintain near-ceiling scores across all metrics. Small communicating arteries remain the most challenging structures, with AComm and PComm exhibiting the lowest Dice scores and highest variance.

 %The M1 segments similarly retain their relatively lower performance with elevated adHD, reflecting not only the intrinsic difficulty of delineating these transitional vessel segments, but also the ambiguity introduced by their definition at vascular branch points (see Figure ~\ref{fig:brain-error-example})

%\begin{figure}
%    \centering
%    \includegraphics[width=1\linewidth]{figure/brain-error-example.pdf}
%    \caption{Voxel-wise comparison between predicted and ground truth vessel segmentations. Green: true positives (TP); red: false positives assigned to a wrong vessel label (FP-label); yellow: false negatives (FN).}
%    \label{fig:brain-error-example}
%\end{figure}

For the TopCoW dataset, consistent findings are observed across all evaluated configurations. The \textit{All labels} configuration trained with Skeleton Recall Loss achieves the best overall performance, with the highest artery mDC of $0.9638 \pm 0.0273$ and tSens of $0.9909 \pm 0.0153$, along with the lowest adHD of $0.0334 \pm 0.0288$ (Tables~\ref{tab:topcow_aggregate_combined}). Per-label results further confirm this trend (Table~\ref{tab:topcow_per_label_dice} and ~\ref{tab:topcow_per_label_skeleton}), where the \textit{All labels} configuration consistently outperforms or matches \textit{All arteries} across most vessel classes, particularly for smaller and more challenging vessels such as AComm and PComm.

\begin{table}[h]
\centering
\caption{Aggregate segmentation performance by vessel mode across label configurations using Dice Loss, Cross-Entropy (CE) Loss, and Skeleton Recall Loss on the TopBrain dataset (mean $\pm$ std).}
\label{tab:aggregate_combined}
\resizebox{\textwidth}{!}{%
\begin{tabular}{lllp{0.1cm} ccc}
\toprule
\textbf{Loss} & \textbf{Configuration} & \textbf{Evaluated vessel} && \textbf{mDC $\uparrow$} & \textbf{tSens $\uparrow$} & \textbf{adHD $\downarrow$} \\
\midrule
\multirow{7}{*}{\textbf{Dice/CE}}
    & \multirow{3}{*}{\textbf{All labels}}
        & All vessel && $0.9116 \pm 0.0296$ & $0.9600 \pm 0.0175$ & $0.1514 \pm 0.0746$ \\
    & & Artery      && $0.8734 \pm 0.0388$ & $0.9387 \pm 0.0257$ & $0.2399 \pm 0.1289$ \\
    & & Vein        && $0.9216 \pm 0.0318$ & $0.9738 \pm 0.0254$ & $0.1057 \pm 0.0482$ \\
\cmidrule(lr){2-7}
    & \textbf{All arteries}
        & Artery     && $0.8847 \pm 0.0407$ & $0.9218 \pm 0.0356$ & $0.2748 \pm 0.1663$ \\
\cmidrule(lr){2-7}
    & \textbf{Names arteries}
        & Artery     && $0.7279 \pm 0.0839$ & $0.7649 \pm 0.0706$ & $1.5167 \pm 0.8808$ \\
\cmidrule(lr){2-7}
    & \textbf{Ignore label}
        & Artery     && $0.8395 \pm 0.0464$ & $0.8475 \pm 0.0462$ & $0.8896 \pm 0.3535$ \\
\midrule
\multirow{4}{*}{\textbf{Skeleton}}
    & \multirow{3}{*}{\textbf{All labels}}
        & All vessel && $0.9381 \pm 0.0217$ & $0.9740 \pm 0.0120$ & $0.1008 \pm 0.0511$ \\
    & & Artery      && $0.9025 \pm 0.0373$ & $0.9477 \pm 0.0228$ & $0.1945 \pm 0.1176$ \\
    & & Vein        && $0.9422 \pm 0.0334$ & $0.9724 \pm 0.0253$ & $0.0758 \pm 0.0435$ \\
\cmidrule(lr){2-7}
    & \textbf{All arteries}
        & Artery     && $0.8859 \pm 0.0427$ & $0.9396 \pm 0.0276$ & $0.2349 \pm 0.1596$ \\
\bottomrule
\end{tabular}%
}
\end{table}

{Table~\ref{tab:arterial_phase} compares multiphase and arterial phase only training under the \textit{All labels} configuration using Skeleton Recall Loss. Multiphase training consistently outperforms arterial phase only training in artery segmentation across both datasets, with mDC improvements of $0.0409$ on TopBrain and $0.0166$ on TopCoW. However, vein segmentation shows a slight decrease of $0.0305$ in mDC compared to arterial phase only training. When compared against a model trained on a subset of TopBrain, the models trained on both the single-phase and multi-phase variants of \textsc{SemanticVessel} outperform the former on two held-out evaluation sets comprising 5 scans from TopBrain and 5 manually annotated arterial phase scans from \textsc{SemanticVessel}  (Tables~\ref{tab:cross_dataset_topbrain}, ~\ref{tab:cross_dataset_semantic},~\ref{tab:per_label_semantic_vessel}, and ~\ref{tab:per_label_topbrain}).} {Additional results are provided in the Appendix, including per-label results with Dice and CE Loss on TopBrain (Table~\ref{tab:per_label_dice}) and full evaluations for TopCoW (Tables~\ref{tab:topcow_aggregate_combined},~\ref{tab:topcow_per_label_dice}, and ~\ref{tab:topcow_per_label_skeleton}).}

\iffalse
\begin{table}[t]
\scriptsize
\centering
\caption{Aggregate segmentation performance by vessel mode across label configurations using Dice Loss and Cross-Entropy (CE) Loss on the TopBrain dataset (mean $\pm$ std).}
\label{tab:aggregate_dice}
\begin{tabular}{ll ccc}
\toprule
& & \textbf{mDC $\uparrow$} & \textbf{tSens $\uparrow$} & \textbf{adHD $\downarrow$} \\
\midrule
\multirow{3}{*}{\textbf{All labels}}
    & All vessel & $0.9116 \pm 0.0296$ & $0.9600 \pm 0.0175$ & $0.1514 \pm 0.0746$ \\
    & Artery        & $0.8734 \pm 0.0388$ & $0.9387 \pm 0.0257$ & $0.2399 \pm 0.1289$ \\
    & Vein          & $0.9216 \pm 0.0318$ & $0.9738 \pm 0.0254$ & $0.1057 \pm 0.0482$ \\
\midrule
\multirow{1}{*}{\textbf{All arteries}}
    & Artery        & $0.8847 \pm 0.0407$ & $0.9218 \pm 0.0356$ & $0.2748 \pm 0.1663$ \\
\midrule
\multirow{1}{*}{\textbf{Names arteries}}
    & Artery        & $0.7279 \pm 0.0839$ & $0.7649 \pm 0.0706$ & $1.5167 \pm 0.8808$ \\
\bottomrule
\end{tabular}
\end{table}
\fi

\begin{table}[h]
    \centering
    \scriptsize
    \setlength{\tabcolsep}{4pt}
    \caption{Per-label segmentation results across label configurations trained with the Skeleton Recall Loss on the TopBrain dataset  (mean $\pm$ std).}
    \label{tab:topbrain_per_label_skeleton}
    \resizebox{\textwidth}{!}{%
    \begin{tabular}{l cc cc cc}
        \toprule
        & \multicolumn{2}{c}{\textbf{mDC $\uparrow$}} 
        & \multicolumn{2}{c}{\textbf{tSens $\uparrow$}} 
        & \multicolumn{2}{c}{\textbf{adHD $\downarrow$}} \\
        \cmidrule(lr){2-3} \cmidrule(lr){4-5} \cmidrule(lr){6-7}
        \textbf{Label} 
        & \textbf{All labels} & \textbf{All arteries}
        & \textbf{All labels} & \textbf{All arteries}
        & \textbf{All labels} & \textbf{All arteries} \\
        \midrule
        ICA (R)      & \textbf{0.7768} $\pm$ 0.0799 & $0.7553 \pm 0.0810$ & \textbf{0.9029} $\pm$ 0.0705 & $0.8989 \pm 0.0704$ & \textbf{0.4256} $\pm$ 0.2191 & $0.4520 \pm 0.2019$ \\
        ICA (L)      & \textbf{0.7848} $\pm$ 0.0687 & $0.7703 \pm 0.0645$ & \textbf{0.9019} $\pm$ 0.1132 & $0.8996 \pm 0.1122$ & \textbf{0.3776} $\pm$ 0.1695 & $0.3955 \pm 0.1512$ \\
        M1 (R)       & $0.6607 \pm 0.1540$ & \textbf{0.6814} $\pm$ 0.1397 & $0.5840 \pm 0.2040$ & \textbf{0.6150} $\pm$ 0.1935 & $1.3845 \pm 0.9958$ & \textbf{1.2027} $\pm$ 0.7869 \\
        M1 (L)       & $0.6748 \pm 0.1463$ & \textbf{0.6844} $\pm$ 0.1506 & $0.5628 \pm 0.2069$ & \textbf{0.5683} $\pm$ 0.2027 & $1.5151 \pm 0.9186$ & \textbf{1.4746} $\pm$ 0.9152 \\
        M2+ (R)      & \textbf{0.9693} $\pm$ 0.0255 & $0.9583 \pm 0.0297$ & \textbf{0.9870} $\pm$ 0.0116 & $0.9807 \pm 0.0197$ & \textbf{0.0271} $\pm$ 0.0248 & $0.0383 \pm 0.0315$ \\
        M2+ (L)      & \textbf{0.9764} $\pm$ 0.0174 & $0.9669 \pm 0.0213$ & \textbf{0.9930} $\pm$ 0.0130 & $0.9895 \pm 0.0133$ & \textbf{0.0170} $\pm$ 0.0168 & $0.0258 \pm 0.0192$ \\
        ACA (R)      & \textbf{0.8923} $\pm$ 0.0914 & $0.8752 \pm 0.0975$ & \textbf{0.9517} $\pm$ 0.0795 & $0.9285 \pm 0.0880$ & \textbf{0.1085} $\pm$ 0.1208 & $0.1602 \pm 0.2327$ \\
        ACA (L)      & $0.8283 \pm 0.0924$ & \textbf{0.8564} $\pm$ 0.0829 & $0.8875 \pm 0.1247$ & \textbf{0.9097} $\pm$ 0.1012 & $0.2525 \pm 0.3390$ & \textbf{0.1872} $\pm$ 0.2718 \\
        AComm        & \textbf{0.6542} $\pm$ 0.2675 & $0.6244 \pm 0.2871$ & \textbf{0.7379} $\pm$ 0.4063 & $0.7359 \pm 0.4272$ & \textbf{0.2715} $\pm$ 0.2451 & $0.4666 \pm 0.8098$ \\
        VA (R)       & \textbf{0.7992} $\pm$ 0.1320 & $0.7620 \pm 0.1299$ & \textbf{0.9268} $\pm$ 0.1281 & $0.9060 \pm 0.1377$ & \textbf{0.4492} $\pm$ 0.6051 & $0.6371 \pm 1.1562$ \\
        VA (L)       & \textbf{0.8605} $\pm$ 0.0749 & $0.8294 \pm 0.0912$ & \textbf{0.9720} $\pm$ 0.0499 & $0.9529 \pm 0.0634$ & \textbf{0.1899} $\pm$ 0.1391 & $0.3368 \pm 0.4004$ \\
        BA           & $0.7561 \pm 0.0851$ & \textbf{0.7600} $\pm$ 0.0863 & $0.9287 \pm 0.0615$ & \textbf{0.9435} $\pm$ 0.0557 & $0.4480 \pm 0.2122$ & \textbf{0.4216} $\pm$ 0.2071 \\
        PCA (R)      & \textbf{0.8729} $\pm$ 0.0877 & $0.8551 \pm 0.1118$ & \textbf{0.8955} $\pm$ 0.0933 & $0.8758 \pm 0.1324$ & \textbf{0.2964} $\pm$ 0.4132 & $0.4201 \pm 0.6317$ \\
        PCA (L)      & \textbf{0.9097} $\pm$ 0.0589 & $0.8923 \pm 0.0790$ & \textbf{0.9368} $\pm$ 0.0629 & $0.9144 \pm 0.0888$ & \textbf{0.1346} $\pm$ 0.1204 & $0.1956 \pm 0.2004$ \\
        PComm (R)    & \textbf{0.6180} $\pm$ 0.3774 & $0.5117 \pm 0.3772$ & \textbf{0.6886} $\pm$ 0.3982 & $0.5565 \pm 0.4134$ & \textbf{0.2899} $\pm$ 0.4676 & $0.7263 \pm 0.9697$ \\
        PComm (L)    & $0.5736 \pm 0.3614$ & \textbf{0.5850} $\pm$ 0.3652 & $0.6033 \pm 0.3899$ & \textbf{0.6433} $\pm$ 0.4057 & \textbf{0.3327} $\pm$ 0.3471 & $0.5488 \pm 0.9391$ \\
        Other artery & \textbf{0.6202} $\pm$ 0.1488 & $0.6199 \pm 0.1360$ & \textbf{0.6569} $\pm$ 0.1331 & $0.6484 \pm 0.1142$ & \textbf{2.1038} $\pm$ 1.8708 & $2.1669 \pm 1.9989$ \\
        Vein         & $0.9422 \pm 0.0334$ & --- & $0.9724 \pm 0.0253$ & --- & $0.0758 \pm 0.0435$ & --- \\
        \bottomrule
    \end{tabular}}
\end{table}
\iffalse
\begin{table}[t]
\scriptsize
\centering
\caption{Aggregate segmentation performance by vessel mode across label configurations using Skeleton Recall Loss on the TopBrain dataset}
\label{tab:aggregate_skeleton}
%\resizebox{\textwidth}{!}{%
\begin{tabular}{ll ccc}
\toprule
& & \textbf{mDC $\uparrow$} & \textbf{tSens $\uparrow$} & \textbf{adHD $\downarrow$} \\
\midrule
\multirow{3}{*}{\textbf{All labels}}
    & Binary vessel & $0.9381 \pm 0.0217$ & $0.9740 \pm 0.0120$ & $0.1008 \pm 0.0511$ \\
    & Artery        & $0.9025 \pm 0.0373$ & $0.9477 \pm 0.0228$ & $0.1945 \pm 0.1176$ \\
    & Vein          & $0.9422 \pm 0.0334$ & $0.9724 \pm 0.0253$ & $0.0758 \pm 0.0435$ \\
\midrule
\multirow{1}{*}{\textbf{All arteries}}
    & Artery        & $XXXXX0.885 \pm 0.048$ & $0.921 \pm 0.033$ & $0.231 \pm 0.157$ \\    
\bottomrule
\end{tabular}%
\end{table}
\fi

\section{Discussion} 

In this paper, we proposed a semi-automatic methodology for fine-grained annotation of arteries in 3D CTA data. By relying on Dynamic 4D CTA, we are able to significantly reduce the annotation burden compared to existing approaches where radiologists need to annotate highly dense 3D volumes from scratch. Our pipeline instead requires only minimal expert intervention. By training our model in three different label configurations, we demonstrated that incorporating unlabeled vessels and complementary vascular structures provides a richer supervisory signal, leading to consistent performance gains. Central to this finding is our explicit treatment of unlabeled vessels. Rather than naively assigning them to the background class, which would introduce conflicting supervisory signal and penalize the model for correctly detecting vessels outside the annotated set, we explicitly handle these regions to avoid false suppression of true vascular predictions. This distinction is particularly important in the context of cerebrovascular anatomy, where the vascular tree is highly interconnected and partially annotated datasets are the norm. By allowing the model to focus its learning on annotated structures without being misled by unannotated ones, our approach ultimately improves its ability to delineate individual arterial branches in fine-grained cerebrovascular segmentation.

\begin{table}[h]
\scriptsize
\centering
\caption{Aggregate segmentation performance with and without left-right coordinate information across label configurations using Dice Loss, Cross-Entropy (CE) Loss on the TopBrain dataset (mean $\pm$ std).}
\label{tab:aggregate_lr}
\resizebox{\columnwidth}{!}{%
\begin{tabular}{llp{0.1cm} cc cc cc}
\toprule
& & & \multicolumn{2}{c}{\textbf{mDC $\uparrow$}} & \multicolumn{2}{c}{\textbf{tSens $\uparrow$}} & \multicolumn{2}{c}{\textbf{adHD $\downarrow$}} \\
\cmidrule(lr){4-5} \cmidrule(lr){6-7} \cmidrule(lr){8-9}
\textbf{Configuration} & \textbf{Evaluated vessel} && \textbf{With LR} & \textbf{Without LR} & \textbf{With LR} & \textbf{Without LR} & \textbf{With LR} & \textbf{Without LR} \\
\midrule
\multirow{3}{*}{\textbf{All labels}}
    & All vessels && $0.9116 \pm 0.0296$ & $0.7876 \pm 0.0556$ & $0.9600 \pm 0.0175$ & $0.8468 \pm 0.0379$ & $0.1514 \pm 0.0746$ & $0.5159 \pm 0.1965$ \\
    & Artery     && $0.8734 \pm 0.0388$ & $0.6453 \pm 0.0499$ & $0.9387 \pm 0.0257$ & $0.7453 \pm 0.0350$ & $0.2399 \pm 0.1289$ & $1.3213 \pm 0.5117$ \\
    & Vein       && $0.9216 \pm 0.0318$ & $0.8545 \pm 0.0713$ & $0.9738 \pm 0.0254$ & $0.8949 \pm 0.0694$ & $0.1057 \pm 0.0482$ & $0.2986 \pm 0.2165$ \\
\bottomrule
\end{tabular}}
\end{table}

\begin{figure}
    \centering
    \includegraphics[width=1\linewidth]{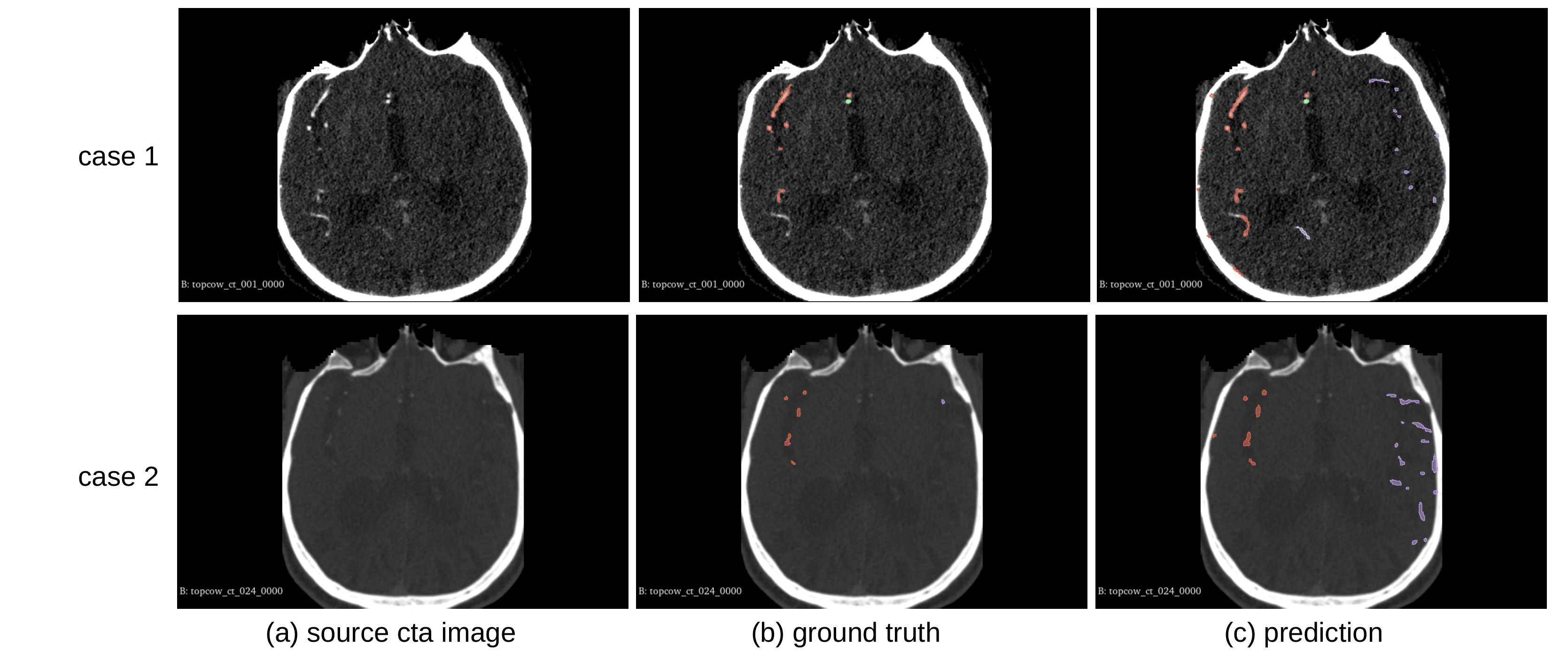}
   \caption{Comparison between ground truth annotations (b) and model predictions (c) on a CTA case from TopBrain. Case 1 (\texttt{topcow\_ct\_001}) The left ICA is occluded. The model predicts additional vessel structures on the left side M2 segment (purple, visible on the right half of column (c)). Case 2 (\texttt{topcow\_ct\_024}) demonstrates poor opacification from the M2 to M3 segment. The model predicts additional vessel structures in the left M3 segment(purple, visible on the right half of column (c)).}
    \label{fig:special-case}
\end{figure}

In addition, we observed that our model predicts more vessel voxels than are present in the ground truth annotations. As illustrated in Figure~\ref{fig:special-case}, expert radiologist verified that these additionally detected structures correspond to true arterial anatomy, suggesting that the model may be recovering anatomically valid vessel structures that were missed or incompletely annotated in the ground truth. These findings indicate that our model is robust and can detect vessels in lower contrast regions, where manual annotation is inherently difficult and prone to under-segmentation.

\begin{figure}
    \centering
    \includegraphics[width=0.7\linewidth]{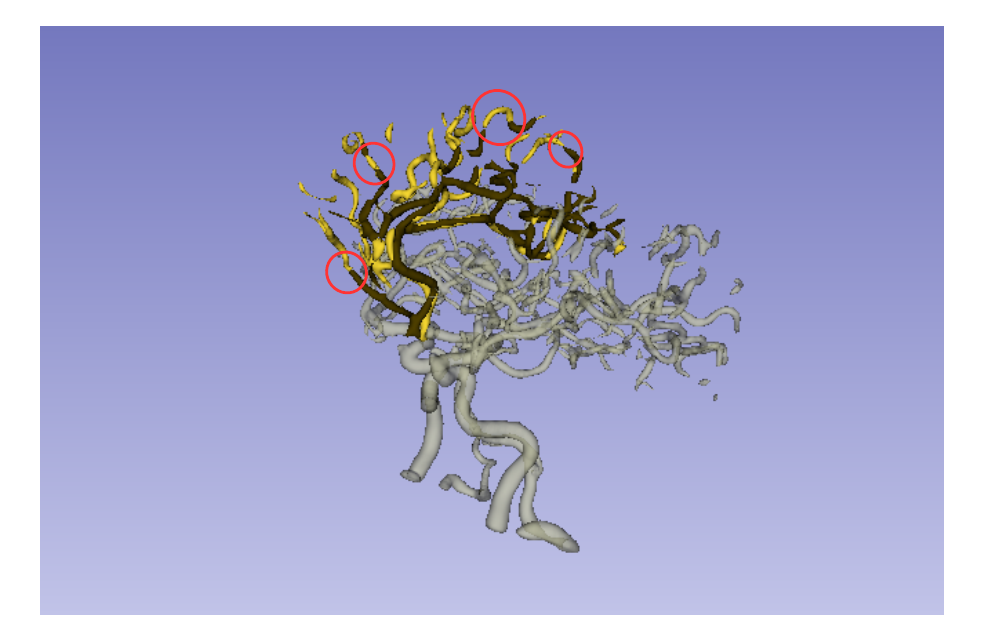}
    \caption{Representative failure cases of ACA segmentation with our best performing model (Skeleton Recall loss). Yellow: right ACA; Brown: left ACA. Red circle: misclassified regions.}
    \label{fig:aca-error}
\end{figure}

Another observation from our experiments is the benefit of incorporating left-right coordinate information during model training. Cerebrovascular structures such as the ICA, M1, ACA, VA, PCA, and PComm are inherently paired, with corresponding branches on both hemispheres sharing similar morphology and spatial relationships. By explicitly encoding left-right coordinate information as an additional input, the model gains awareness of hemispheric context, enabling it to better distinguish between left and right counterparts and exploit their structural symmetry as a complementary supervisory signal. Table~\ref{tab:aggregate_lr} presents the segmentation performance comparison between models trained with and without left-right coordinate information. However, certain structures, such as the ACA  especially in the A2 and more distal segments, remain challenging to separate due to their complex anatomy and close proximity to the midline of the brain (see Figure ~\ref{fig:aca-error}).

%\emph{This is probably the most important section of your paper!  This
%  is where you tell us how your work advances our understanding of
%  machine learning and healthcare.}  Discuss both technical and
%clinical implications, as appropriate\cite{xyz19}.

\paragraph{Limitations}

Although \dvplus    is the largest dataset for fine-grained brain artery segmentation in CTA to date, key limitations have to be considered:

\begin{itemize}
    \item Due to the presence of symmetrical structures on the left and right sides of the brain, we are forced to use an additional channel encoding laterality, which itself is derived from a brain mask produced by an off-the-shelf brain segmentation tool; in cases where the brain segmentation fails (\textit{e.g.}, skull-stripped scans) performance can be expected to go down significantly.
    \item Our annotation scheme is more limited than that of TopBrain, which covers 34 arterial and 6 venous structures compared to the 21 arterial structures and a single venous label in our dataset; the lack of specificity could make models trained on \dvplus  less useful in certain scenarios, such as grounded vision-language modeling.
    \item Our dataset has no coverage of occluded arteries or anatomical variants that are somewhat common in real-world scenarios; for instance, 9 out of 25 CTA scans in TopBrain are from patients without a right posterior communicating artery. Although the nnUNet trained on our data often does not make predictions for CTA with missing arteries, a comprehensive analysis with a larger number of patients with anatomical variants will be necessary.
    \item Since our annotations are not fully connected in all cases, predictions made by the nnUNet model sometimes display gaps; this limits the usability of our dataset for dynamic fluids simulations, which require fully accurate vessel trees and are critical tools in diagnosing patient risk under certain cerebrovascular diseases.
    \item {The region growing step in our annotation pipeline involves heuristic parameters that may require case-specific tuning when applied to data from different scanners or acquisition protocols. While our interactive GUI tool allows radiologists to efficiently review and correct suboptimal segmentations, this represents an inherent limitation of the heuristic pipeline design that may increase the annotation burden for institutions with different imaging characteristics.}
\end{itemize}

%Explain when your approach may not apply, or things you could not check.  \emph{Discussing limitations is essential.  Both ACs and   reviewers have been advised to be skeptical of any work that does   not consider limitations.}

% ACKNOWLEDGEMENTS ONLY GO IN THE CAMERA-READY, NOT THE SUBMISSION
% \acks{Many thanks to all collaborators and funders!}

\acks{The authors acknowledge the financial support provided by NIH grant 1R01LM013891-01A1. Ceballos Arroyo, A. received funding from Colombia’s Minciencias and Fulbright under the Fulbright Minciencias 2021 program. This work was partially supported by the DeltaAI system at the National Center for Supercomputing Applications through allocation CIS260200 from the Advanced Cyberinfrastructure Coordination Ecosystem: Services \& Support (ACCESS) program, which is supported by National Science Foundation grants \#2138259, \#2138286, \#2138307, \#2137603, and \#2138296.}

%Do NOT change font size of references or modify the bibliography style
%\bibliography{SemanticVesselseg}

\begin{thebibliography}{30}
\providecommand{\natexlab}[1]{#1}
\providecommand{\url}[1]{\texttt{#1}}
\expandafter\ifx\csname urlstyle\endcsname\relax
  \providecommand{\doi}[1]{doi: #1}\else
  \providecommand{\doi}{doi: \begingroup \urlstyle{rm}\Url}\fi

\bibitem[Bash et~al.(2005)Bash, Villablanca, Jahan, Duckwiler, Tillis, Kidwell, Saver, and Sayre]{bash_intracranial_2005}
Suzie Bash, J.~Pablo Villablanca, Reza Jahan, Gary Duckwiler, Monica Tillis, Chelsea Kidwell, Jeffrey Saver, and James Sayre.
\newblock Intracranial vascular stenosis and occlusive disease: evaluation with {CT} angiography, {MR} angiography, and digital subtraction angiography.
\newblock \emph{AJNR Am J Neuroradiol}, 26\penalty0 (5):\penalty0 1012--1021, May 2005.
\newblock ISSN 0195-6108.

\bibitem[Ceballos-Arroyo et~al.(2026)]{ceballos2026dynavessels}
Alberto~M. Ceballos-Arroyo et~al.
\newblock Robust automatic brain vessel segmentation in {3D CTA} scans using dynamic {4D-CTA} data.
\newblock \emph{arXiv preprint arXiv:2602.00391}, 2026.
\newblock URL \url{https://arxiv.org/abs/2602.00391}.

\bibitem[Chen et~al.(2019)Chen, Mossa-Basha, Sun, Hippe, Balu, Yuan, Pimentel, Hatsukami, Hwang, and Yuan]{chen_quantification_2019}
Li~Chen, Mahmud Mossa-Basha, Jie Sun, Daniel~S. Hippe, Niranjan Balu, Quan Yuan, Kristi Pimentel, Thomas~S. Hatsukami, Jenq-Neng Hwang, and Chun Yuan.
\newblock Quantification of morphometry and intensity features of intracranial arteries from {3D} {TOF} {MRA} using the intracranial artery feature extraction ({iCafe}): {A} reproducibility study.
\newblock \emph{Magn Reson Imaging}, 57:\penalty0 293--302, April 2019.
\newblock ISSN 1873-5894.
\newblock \doi{10.1016/j.mri.2018.12.007}.

\bibitem[Chen et~al.(2024)Chen, You, Zhang, Ye, Feng, Lu, Lv, Tang, Wei, Gui, et~al.]{chen2024cta}
Ting Chen, Wei You, Liyuan Zhang, Wanxing Ye, Junqiang Feng, Jing Lu, Jian Lv, Yudi Tang, Dachao Wei, Siming Gui, et~al.
\newblock Automated anatomical labeling of the intracranial arteries via deep learning in computed tomography angiography.
\newblock \emph{Frontiers in Physiology}, 14:\penalty0 1310357, 2024.
\newblock \doi{10.3389/fphys.2023.1310357}.

\bibitem[Chen et~al.(2018)Chen, Liu, Tong, Dong, Ma, Xu, and Yang]{chen_meta-analysis_2018}
Xiaodan Chen, Yun Liu, Huazhang Tong, Yonghai Dong, Dongyang Ma, Lei Xu, and Cheng Yang.
\newblock Meta-analysis of computed tomography angiography versus magnetic resonance angiography for intracranial aneurysm.
\newblock \emph{Medicine (Baltimore)}, 97\penalty0 (20):\penalty0 e10771, May 2018.
\newblock ISSN 0025-7974.
\newblock \doi{10.1097/MD.0000000000010771}.
\newblock URL \url{https://pmc.ncbi.nlm.nih.gov/articles/PMC5976319/}.

\bibitem[Falcetta et~al.(2025)Falcetta, Marciano, Yang, Cleary, Legris, Rizzaro, Pitsiorlas, Chaptoukaev, Lemasson, Menze, and Zuluaga]{FalDan_VesselVerse_MICCAI2025}
Daniele Falcetta, Vincenzo Marciano, Kaiyuan Yang, Jon Cleary, Loïc Legris, Massimiliano~Domenico Rizzaro, Ioannis Pitsiorlas, Hava Chaptoukaev, Benjamin Lemasson, Bjoern Menze, and Maria~A. Zuluaga.
\newblock { VesselVerse: A Dataset and Collaborative Framework for Vessel Annotation }.
\newblock In \emph{proceedings of Medical Image Computing and Computer Assisted Intervention -- MICCAI 2025}, volume LNCS 15972. Springer Nature Switzerland, September 2025.

\bibitem[Falcetta et~al.(2026)Falcetta, Canas, Suppa, Pentassuglia, Cleary, Modat, Ourselin, and Zuluaga]{falcetta2026automatedframeworklargescalegraphbased}
Daniele Falcetta, Liane~S. Canas, Lorenzo Suppa, Matteo Pentassuglia, Jon Cleary, Marc Modat, Sébastien Ourselin, and Maria~A. Zuluaga.
\newblock An automated framework for large-scale graph-based cerebrovascular analysis, 2026.
\newblock URL \url{https://arxiv.org/abs/2512.03869}.

\bibitem[Fu et~al.(2020)Fu, Wei, Zhang, Yu, Xiao, Rong, Shan, Li, Zhao, Liao, Yang, Li, Chen, Wang, and Lu]{fu_rapid_2020}
Fan Fu, Jianyong Wei, Miao Zhang, Fan Yu, Yueting Xiao, Dongdong Rong, Yi~Shan, Yan Li, Cheng Zhao, Fangzhou Liao, Zhenghan Yang, Yuehua Li, Yingmin Chen, Ximing Wang, and Jie Lu.
\newblock Rapid vessel segmentation and reconstruction of head and neck angiograms using {3D} convolutional neural network.
\newblock \emph{Nat Commun}, 11\penalty0 (1):\penalty0 4829, September 2020.
\newblock ISSN 2041-1723.
\newblock \doi{10.1038/s41467-020-18606-2}.
\newblock URL \url{https://www.nature.com/articles/s41467-020-18606-2}.

\bibitem[Hamadache et~al.(2026)Hamadache, Lisazo, Yalcin, Lal-Trehan~Estrada, Abramova, Casamitjana, Oliver, and Llad{\'o}]{hamadache2026topology}
Rachika~E. Hamadache, Clara Lisazo, Cansu Yalcin, Uma~M. Lal-Trehan~Estrada, Valeriia Abramova, Adri{\`a} Casamitjana, Arnau Oliver, and Xavier Llad{\'o}.
\newblock Topology-aware multiclass segmentation of the {Circle of Willis} from {MRA} and {CTA} images.
\newblock \emph{Computers in Biology and Medicine}, 204:\penalty0 111516, 2026.
\newblock \doi{10.1016/j.compbiomed.2026.111516}.

\bibitem[Hilbert et~al.(2020)Hilbert, Madai, Akay, Aydin, Behland, Sobesky, Galinovic, Khalil, Taha, Wuerfel, Dusek, Niendorf, Fiebach, Frey, and Livne]{hilbert_brave-net_2020}
Adam Hilbert, Vince~I. Madai, Ela~M. Akay, Orhun~U. Aydin, Jonas Behland, Jan Sobesky, Ivana Galinovic, Ahmed~A. Khalil, Abdel~A. Taha, Jens Wuerfel, Petr Dusek, Thoralf Niendorf, Jochen~B. Fiebach, Dietmar Frey, and Michelle Livne.
\newblock {BRAVE}-{NET}: {Fully} {Automated} {Arterial} {Brain} {Vessel} {Segmentation} in {Patients} {With} {Cerebrovascular} {Disease}.
\newblock \emph{Front. Artif. Intell.}, 3, September 2020.
\newblock ISSN 2624-8212.
\newblock \doi{10.3389/frai.2020.552258}.
\newblock URL \url{https://www.frontiersin.org/journals/artificial-intelligence/articles/10.3389/frai.2020.552258/full}.

\bibitem[Isensee et~al.(2021)Isensee, Jaeger, Kohl, Petersen, and Maier-Hein]{isensee_nnu-net_2021}
Fabian Isensee, Paul~F. Jaeger, Simon A.~A. Kohl, Jens Petersen, and Klaus~H. Maier-Hein.
\newblock {nnU}-{Net}: a self-configuring method for deep learning-based biomedical image segmentation.
\newblock \emph{Nat Methods}, 18\penalty0 (2):\penalty0 203--211, February 2021.
\newblock ISSN 1548-7105.
\newblock \doi{10.1038/s41592-020-01008-z}.
\newblock URL \url{https://www.nature.com/articles/s41592-020-01008-z}.

\bibitem[Kikinis et~al.(2014)Kikinis, Pieper, and Vosburgh]{Kikinis2014}
Ron Kikinis, Steve~D. Pieper, and Kirby~G. Vosburgh.
\newblock \emph{3D Slicer: A Platform for Subject-Specific Image Analysis, Visualization, and Clinical Support}, pages 277--289.
\newblock Springer New York, New York, NY, 2014.
\newblock ISBN 978-1-4614-7657-3.
\newblock URL \url{https://doi.org/10.1007/978-1-4614-7657-3_19}.

\bibitem[Kirchhoff et~al.(2024)Kirchhoff, Rokuss, Roy, Kovacs, Ulrich, Wald, Zenk, Vollmuth, Kleesiek, Isensee, and Maier-Hein]{kirchhoff_skeleton_2024}
Yannick Kirchhoff, Maximilian~R. Rokuss, Saikat Roy, Balint Kovacs, Constantin Ulrich, Tassilo Wald, Maximilian Zenk, Philipp Vollmuth, Jens Kleesiek, Fabian Isensee, and Klaus Maier-Hein.
\newblock Skeleton {Recall} {Loss} for {Connectivity} {Conserving} and {Resource} {Efficient} {Segmentation} of {Thin} {Tubular} {Structures}.
\newblock In Aleš Leonardis, Elisa Ricci, Stefan Roth, Olga Russakovsky, Torsten Sattler, and Gül Varol, editors, \emph{Computer {Vision} – {ECCV} 2024}, pages 218--234, Cham, 2024. Springer Nature Switzerland.
\newblock ISBN 978-3-031-72980-5.
\newblock \doi{10.1007/978-3-031-72980-5_13}.

\bibitem[Ma et~al.(2024)Ma, He, Li, Han, You, and Wang]{ma_segment_2024}
Jun Ma, Yuting He, Feifei Li, Lin Han, Chenyu You, and Bo~Wang.
\newblock Segment anything in medical images.
\newblock \emph{Nat Commun}, 15\penalty0 (1):\penalty0 654, January 2024.
\newblock ISSN 2041-1723.
\newblock \doi{10.1038/s41467-024-44824-z}.
\newblock URL \url{https://www.nature.com/articles/s41467-024-44824-z}.

\bibitem[Meijs et~al.(2017)Meijs, Patel, van~de Leemput, Prokop, van Dijk, de~Leeuw, Meijer, van Ginneken, and Manniesing]{meijs_robust_2017}
Midas Meijs, Ajay Patel, Sil~C. van~de Leemput, Mathias Prokop, Ewoud~J. van Dijk, Frank-Erik de~Leeuw, Frederick J.~A. Meijer, Bram van Ginneken, and Rashindra Manniesing.
\newblock Robust {Segmentation} of the {Full} {Cerebral} {Vasculature} in {4D} {CT} of {Suspected} {Stroke} {Patients}.
\newblock \emph{Sci Rep}, 7\penalty0 (1):\penalty0 15622, November 2017.
\newblock ISSN 2045-2322.
\newblock \doi{10.1038/s41598-017-15617-w}.
\newblock URL \url{https://www.nature.com/articles/s41598-017-15617-w}.

\bibitem[Min et~al.(2024)Min, Li, Jia, Li, and Nie]{min_automated_2024}
Yuqin Min, Jing Li, Shouqiang Jia, Yuehua Li, and Shengdong Nie.
\newblock Automated {Cerebrovascular} {Segmentation} and {Visualization} of {Intracranial} {Time}-of-{Flight} {Magnetic} {Resonance} {Angiography} {Based} on {Deep} {Learning}.
\newblock \emph{J Imaging Inform Med}, 38\penalty0 (2):\penalty0 703--716, August 2024.
\newblock ISSN 2948-2925.
\newblock \doi{10.1007/s10278-024-01215-6}.
\newblock URL \url{https://pmc.ncbi.nlm.nih.gov/articles/PMC11950609/}.

\bibitem[Patel et~al.(2023)Patel, Patel, Veeturi, Shah, Waqas, Monteiro, Baig, Pinter, Levy, Siddiqui, and Tutino]{patel_evaluating_2023}
Tatsat~R. Patel, Aakash Patel, Sricharan~S. Veeturi, Munjal Shah, Muhammad Waqas, Andre Monteiro, Ammad~A. Baig, Nandor Pinter, Elad~I. Levy, Adnan~H. Siddiqui, and Vincent~M. Tutino.
\newblock Evaluating a {3D} deep learning pipeline for cerebral vessel and intracranial aneurysm segmentation from computed tomography angiography-digital subtraction angiography image pairs.
\newblock \emph{Neurosurg Focus}, 54\penalty0 (6):\penalty0 E13, June 2023.
\newblock ISSN 1092-0684.
\newblock \doi{10.3171/2023.3.FOCUS2374}.

\bibitem[Phillips and Bubash(2002)]{phillips_ct_2002}
C.~Douglas Phillips and Lori~A. Bubash.
\newblock {CT} angiography and {MR} angiography in the evaluation of extracranial carotid vascular disease.
\newblock \emph{Radiol Clin North Am}, 40\penalty0 (4):\penalty0 783--798, July 2002.
\newblock ISSN 0033-8389.
\newblock \doi{10.1016/s0033-8389(02)00017-9}.

\bibitem[Rist et~al.(2023)Rist, Taubmann, Thamm, Ditt, S{\"u}hling, and Maier]{rist2023bifurcation}
Leonhard Rist, Oliver Taubmann, Florian Thamm, Hendrik Ditt, Michael S{\"u}hling, and Andreas Maier.
\newblock Bifurcation matching for consistent cerebral vessel labeling in {CTA} of stroke patients.
\newblock \emph{International Journal of Computer Assisted Radiology and Surgery}, 18\penalty0 (3):\penalty0 509--516, 2023.
\newblock \doi{10.1007/s11548-022-02750-9}.

\bibitem[Saluja et~al.(2025)Saluja, Cihangir, Deng, Paetzold, Liu, and Sabuncu]{saluja_backsplit_2025}
Rachit Saluja, Asli Cihangir, Ruining Deng, Johannes~C. Paetzold, Fengbei Liu, and Mert~R. Sabuncu.
\newblock {BackSplit}: {The} {Importance} of {Sub}-dividing the {Background} in {Biomedical} {Lesion} {Segmentation}, November 2025.
\newblock URL \url{http://arxiv.org/abs/2511.19394}.
\newblock arXiv:2511.19394 [cs].

\bibitem[Shit et~al.(2021)Shit, Paetzold, Sekuboyina, Ezhov, Unger, Zhylka, Pluim, Bauer, and Menze]{shit_cldice_2021}
Suprosanna Shit, Johannes~C. Paetzold, Anjany Sekuboyina, Ivan Ezhov, Alexander Unger, Andrey Zhylka, Josien P.~W. Pluim, Ulrich Bauer, and Bjoern~H. Menze.
\newblock {clDice} - a {Novel} {Topology}-{Preserving} {Loss} {Function} for {Tubular} {Structure} {Segmentation}.
\newblock In \emph{2021 {IEEE}/{CVF} {Conference} on {Computer} {Vision} and {Pattern} {Recognition} ({CVPR})}, pages 16555--16564, Nashville, TN, USA, June 2021. IEEE.
\newblock ISBN 978-1-66544-509-2.
\newblock URL \url{https://ieeexplore.ieee.org/document/9578225/}.

\bibitem[Talou et~al.(2021)Talou, Safaei, Hunter, and Blanco]{talou_adaptive_2021}
Gonzalo Daniel~Maso Talou, Soroush Safaei, Peter~John Hunter, and Pablo~Javier Blanco.
\newblock Adaptive constrained constructive optimisation for complex vascularisation processes.
\newblock \emph{Scientific Reports}, 11\penalty0 (1):\penalty0 6180, March 2021.
\newblock ISSN 2045-2322.
\newblock URL \url{https://www.nature.com/articles/s41598-021-85434-9}.
\newblock Publisher: Nature Publishing Group.

\bibitem[Thamm et~al.(2022)]{thamm2022labeling}
Florian Thamm et~al.
\newblock An algorithm for the labeling and interactive visualization of the cerebrovascular system of ischemic strokes.
\newblock \emph{Biomedical Physics \& Engineering Express}, 8\penalty0 (6):\penalty0 065016, 2022.
\newblock \doi{10.1088/2057-1976/ac9107}.

\bibitem[Tustison et~al.(2021)Tustison, Cook, Holbrook, Johnson, Muschelli, Devenyi, Duda, Das, Cullen, Gillen, Yassa, Stone, Gee, and Avants]{tustison_antsx_2021}
Nicholas~J. Tustison, Philip~A. Cook, Andrew~J. Holbrook, Hans~J. Johnson, John Muschelli, Gabriel~A. Devenyi, Jeffrey~T. Duda, Sandhitsu~R. Das, Nicholas~C. Cullen, Daniel~L. Gillen, Michael~A. Yassa, James~R. Stone, James~C. Gee, and Brian~B. Avants.
\newblock The {ANTsX} ecosystem for quantitative biological and medical imaging.
\newblock \emph{Scientific Reports}, 11\penalty0 (1):\penalty0 9068, April 2021.
\newblock ISSN 2045-2322.
\newblock \doi{10.1038/s41598-021-87564-6}.
\newblock URL \url{https://doi.org/10.1038/s41598-021-87564-6}.

\bibitem[van Voorst et~al.(2026)van Voorst, Su, Konduri, Majoie, Roos, Emmer, Marquering, de~Vos, Caan, Išgum, and {MR CLEAN Registry collaborators}]{van_voorst_deep_2026}
Henk van Voorst, Jiahang Su, Praneeta~R. Konduri, Charles B. L.~M. Majoie, Yvo B. W. E.~M. Roos, Bart~J. Emmer, Henk~A. Marquering, Bob~D. de~Vos, Matthan W.~A. Caan, Ivana Išgum, and {MR CLEAN Registry collaborators}.
\newblock Deep generative models for vessel segmentation in {CT} angiography of the brain.
\newblock \emph{Comput Biol Med}, 202:\penalty0 111432, February 2026.
\newblock ISSN 1879-0534.
\newblock \doi{10.1016/j.compbiomed.2025.111432}.

\bibitem[Wasserthal et~al.(2023)Wasserthal, Breit, Meyer, Pradella, Hinck, Sauter, Heye, Boll, Cyriac, Yang, Bach, and Segeroth]{wasserthal_totalsegmentator_2023}
Jakob Wasserthal, Hanns-Christian Breit, Manfred~T. Meyer, Maurice Pradella, Daniel Hinck, Alexander~W. Sauter, Tobias Heye, Daniel~T. Boll, Joshy Cyriac, Shan Yang, Michael Bach, and Martin Segeroth.
\newblock {TotalSegmentator}: {Robust} {Segmentation} of 104 {Anatomic} {Structures} in {CT} {Images}.
\newblock \emph{Radiology: Artificial Intelligence}, 5\penalty0 (5):\penalty0 e230024, September 2023.
\newblock \doi{10.1148/ryai.230024}.
\newblock URL \url{https://pubs.rsna.org/doi/full/10.1148/ryai.230024}.

\bibitem[Yang et~al.(2025)Yang, Musio, Ma, Juchler, et~al.]{yang2025topcow}
Kaiyuan Yang, Fabio Musio, Yihui Ma, Norman Juchler, et~al.
\newblock Benchmarking the {CoW} with the {TopCoW} challenge: Topology-aware anatomical segmentation of the circle of {Willis} for {CTA} and {MRA}, 2025.

\bibitem[Yang et~al.(2026)Yang, Shi, Huang, Musio, Baazaoui, Aydin, Hilbert, Hamadache, Yalcin, Zhang, Falcetta, Rosa, Shit, Prabhakar, Wittmann, Rokuss, Kirchhoff, Al-Maskari, Höher, Juchler, Casamitjana, Cleary, Schmick, Baumgartner, Deseö, Vandans, Lee, Oh, LaBella, Mazher, Niederer, Qayyum, Liu, Chen, Kim, Asawalertsak, Kim, Shin, Park, Kikuchi, Zhang, Liu, Cui, Qiu, Verschuur, Zhang, Schaaf, Su, Tax, Yamagishi, Letchumanan, Hanaoka, González, Tiberi, Lisazo, Abramova, Estrada, Lathulerie, Díaz, Oliver, Zhang, Bai, Maier-Hein, Gu, Lladó, Zuluaga, Frey, Paetzold, Hirsch, Wegener, Ma, and Menze]{yang_topbrain_2026}
Kaiyuan Yang, Pengcheng Shi, Houjing Huang, Fabio Musio, Hakim Baazaoui, Orhun~Utku Aydin, Adam Hilbert, Rachika~E. Hamadache, Cansu Yalcin, Minghui Zhang, Daniele Falcetta, Ezequiel de~la Rosa, Suprosanna Shit, Chinmay Prabhakar, Bastian Wittmann, Maximilian~R. Rokuss, Yannick Kirchhoff, Rami Al-Maskari, Luciano Höher, Norman Juchler, Adrià Casamitjana, Jon Cleary, Anton Schmick, Philipp Baumgartner, Julian Deseö, Olafs Vandans, Dahye Lee, Kwanseok Oh, Dominic LaBella, Moona Mazher, Steven~A. Niederer, Abdul Qayyum, Yaoyu Liu, Junqiang Chen, Wooseung Kim, Napasara Asawalertsak, Minjae Kim, Dongho Shin, Sung-Hong Park, Shunsuke Kikuchi, Yaqing Zhang, Jialu Liu, Yue Cui, Yuchen Qiu, Anouk Verschuur, Jiaxin Zhang, Irene van~der Schaaf, Ruisheng Su, Chantal M.~W. Tax, Yosuke Yamagishi, Nishta Letchumanan, Shouhei Hanaoka, Jesús González, Riccardo Tiberi, Clara Lisazo, Valeriia Abramova, Uma Maria Lal-Trehan Estrada, Agustin~Cartaya Lathulerie, Micaela~Rivas Díaz, Arnau Oliver, Menghan Zhang, Zhiqiang Bai,
  Klaus Maier-Hein, Yun Gu, Xavier Lladó, Maria~A. Zuluaga, Dietmar Frey, Johannes~C. Paetzold, Sven Hirsch, Susanne Wegener, Yihui Ma, and Bjoern Menze.
\newblock {TopBrain} {Segmentation} {Challenge} for {Whole} {Brain} {Vessel} {Anatomy}, May 2026.
\newblock URL \url{https://www.medrxiv.org/content/10.64898/2026.05.28.26354312v1}.
\newblock ISSN: 3067-2007 Pages: 2026.05.28.26354312.

\bibitem[Zhang et~al.(2024)]{zhang2024topcow24}
Minghui Zhang et~al.
\newblock Topology-aware exploration of {Circle of Willis} for {CTA} and {MRA}: Segmentation, detection, and classification.
\newblock \emph{arXiv preprint arXiv:2410.15614}, 2024.
\newblock URL \url{https://arxiv.org/abs/2410.15614}.

\bibitem[Zhou et~al.(2024)Zhou, Wu, Luo, and Zhou]{zhou_deep_2024}
Langtao Zhou, Huiting Wu, Guanghua Luo, and Hong Zhou.
\newblock Deep learning-based {3D} cerebrovascular segmentation workflow on bright and black blood sequences magnetic resonance angiography.
\newblock \emph{Insights Imaging}, 15\penalty0 (1):\penalty0 81, March 2024.
\newblock ISSN 1869-4101.
\newblock \doi{10.1186/s13244-024-01657-0}.

\end{thebibliography}

\newpage
\appendix
\section*{Appendix}
\renewcommand{\thetable}{A\arabic{table}}
\setcounter{table}{0}

\begin{table}[H]
\centering
\small
\caption{Mapping of TopBrain and TopCoW vessel labels to the classes in \textsc{SemanticVessel}.}
\label{tab:topbrain_topcow_semanticvessel}
\renewcommand{\arraystretch}{0.9}
\begin{tabular}{p{4.5cm}p{4.5cm}p{4.5cm}}
\toprule
\textbf{SemanticVessel} & \textbf{TopBrain} & \textbf{TopCoW} \\
\midrule
ICA (R), ICA (L)     & R-ICA, L-ICA                   & R-ICA, L-ICA \\
M1 (R), M1 (L)       & R-M1, L-M1                     & R-MCA, L-MCA \\
M2+ (R), M2+ (L)     & R-M2, R-M3, L-M2, L-M3        & --- \\
ACA (R)              & R-A1A2, R-A3                   & R-ACA \\
ACA (L)              & L-A1A2, L-A3                   & L-ACA \\
PCA (R)              & R-P1P2, R-P3P4                 & R-PCA \\
PCA (L)              & L-P1P2, L-P3P4                 & L-PCA \\
PComm (R), PComm (L) & R-Pcom, L-Pcom                 & R-Pcom, L-Pcom \\
AComm                & Acom                           & Acom \\
BA                   & BA                             & BA \\
VA (R), VA (L)       & R-VA, L-VA                     & --- \\
\midrule
Other artery         & \parbox[t]{4.5cm}{\raggedright 3rd-A2, 3rd-A3, R-SCA, L-SCA, R-AICA, L-AICA, R-PICA, L-PICA, R-AChA, L-AChA, R-OA, L-OA} & 3rd-A2 \\
\midrule
Vein                 & VoG, StS, ICVs, R-BVR, L-BVR, SSS & --- \\
\bottomrule
\end{tabular}
\end{table}

\begin{table}[H]
    \centering
    \scriptsize
    \setlength{\tabcolsep}{4pt}
    \caption{Per-label segmentation results across label configurations trained with Dice Loss and Cross-Entropy (CE) Loss (mean $\pm$ std) on the TopBrain dataset. }
    \label{tab:per_label_dice}
    \resizebox{\textwidth}{!}{%
    \begin{tabular}{l ccc ccc ccc}
        \toprule
        & \multicolumn{3}{c}{\textbf{mDC $\uparrow$}} 
        & \multicolumn{3}{c}{\textbf{tSens $\uparrow$}} 
        & \multicolumn{3}{c}{\textbf{adHD $\downarrow$}} \\
        \cmidrule(lr){2-4} \cmidrule(lr){5-7} \cmidrule(lr){8-10}
        \textbf{Label} 
        & \textbf{All labels} & \textbf{All arteries} & \textbf{Named arteries}
        & \textbf{All labels} & \textbf{All arteries} & \textbf{Named arteries}
        & \textbf{All labels} & \textbf{All arteries} & \textbf{Named arteries} \\
        \midrule
        ICA (R)      & \textbf{0.7668 $\pm$ 0.0745} & 0.7593 $\pm$ 0.1461 & 0.6849 $\pm$ 0.1037 & \textbf{0.9338 $\pm$ 0.0585} & 0.9080 $\pm$ 0.1443 & 0.8961 $\pm$ 0.1254 & \textbf{0.4091 $\pm$ 0.1777} & 0.4476 $\pm$ 0.3040 & 0.6087 $\pm$ 0.5091 \\
        ICA (L)      & 0.7824 $\pm$ 0.0699 & \textbf{0.7863 $\pm$ 0.0820} & 0.6776 $\pm$ 0.1435 & 0.9114 $\pm$ 0.0969 & \textbf{0.9121 $\pm$ 0.1057} & 0.8634 $\pm$ 0.1848 & 0.3511 $\pm$ 0.1454 & \textbf{0.3461 $\pm$ 0.1528} & 0.5555 $\pm$ 0.3857 \\
        M1 (R)       & 0.6438 $\pm$ 0.1382 & \textbf{0.6673 $\pm$ 0.1505} & 0.5971 $\pm$ 0.1256 & 0.5731 $\pm$ 0.2057 & \textbf{0.5875 $\pm$ 0.2009} & 0.5657 $\pm$ 0.2076 & 1.3758 $\pm$ 0.8946 & \textbf{1.2982 $\pm$ 0.9273} & 1.4593 $\pm$ 0.9582 \\
        M1 (L)       & 0.6351 $\pm$ 0.1490 & \textbf{0.6568 $\pm$ 0.1591} & 0.6520 $\pm$ 0.1477 & 0.5164 $\pm$ 0.2156 & 0.5497 $\pm$ 0.2192 & \textbf{0.5736 $\pm$ 0.2033} & 1.7690 $\pm$ 1.0639 & 1.5824 $\pm$ 1.0622 & \textbf{1.4142 $\pm$ 0.9039} \\
        M2+ (R)      & \textbf{0.9407 $\pm$ 0.0420} & 0.9165 $\pm$ 0.0463 & 0.8836 $\pm$ 0.0588 & \textbf{0.9768 $\pm$ 0.0253} & 0.9549 $\pm$ 0.0353 & 0.9214 $\pm$ 0.0514 & \textbf{0.0514 $\pm$ 0.0444} & 0.1181 $\pm$ 0.1068 & 0.1767 $\pm$ 0.1254 \\
        M2+ (L)      & \textbf{0.9592 $\pm$ 0.0260} & 0.9438 $\pm$ 0.0300 & 0.9203 $\pm$ 0.0515 & \textbf{0.9901 $\pm$ 0.0108} & 0.9711 $\pm$ 0.0244 & 0.9601 $\pm$ 0.0441 & \textbf{0.0285 $\pm$ 0.0198} & 0.0673 $\pm$ 0.0523 & 0.0955 $\pm$ 0.1080 \\
        ACA (R)      & 0.8830 $\pm$ 0.0883 & \textbf{0.8831 $\pm$ 0.0958} & 0.8698 $\pm$ 0.0527 & 0.9493 $\pm$ 0.0747 & 0.9339 $\pm$ 0.0813 & \textbf{0.9681 $\pm$ 0.0440} & 0.1282 $\pm$ 0.1435 & 0.1504 $\pm$ 0.1668 & \textbf{0.1222 $\pm$ 0.0792} \\
        ACA (L)      & 0.8137 $\pm$ 0.1090 & \textbf{0.8689 $\pm$ 0.0812} & 0.7734 $\pm$ 0.1106 & 0.8868 $\pm$ 0.1244 & \textbf{0.9290 $\pm$ 0.0945} & 0.8849 $\pm$ 0.1077 & 0.2425 $\pm$ 0.2312 & \textbf{0.1589 $\pm$ 0.1925} & 0.3111 $\pm$ 0.2817 \\
        AComm        & 0.4757 $\pm$ 0.3729 & \textbf{0.5342 $\pm$ 0.3544} & 0.3122 $\pm$ 0.3214 & 0.5286 $\pm$ 0.4752 & \textbf{0.5835 $\pm$ 0.4729} & 0.4929 $\pm$ 0.4792 & 0.5396 $\pm$ 0.9225 & \textbf{0.4809 $\pm$ 0.7639} & 0.4778 $\pm$ 0.3539 \\
        VA (R)       & 0.7246 $\pm$ 0.1513 & \textbf{0.8306 $\pm$ 0.0744} & 0.4415 $\pm$ 0.2579 & 0.8776 $\pm$ 0.1906 & \textbf{0.9480 $\pm$ 0.1106} & 0.5305 $\pm$ 0.3381 & 1.2630 $\pm$ 3.8663 & \textbf{0.3165 $\pm$ 0.4529} & 3.4846 $\pm$ 3.8562 \\
        VA (L)       & 0.8148 $\pm$ 0.0585 & \textbf{0.8443 $\pm$ 0.0653} & 0.6175 $\pm$ 0.2328 & \textbf{0.9601 $\pm$ 0.0556} & 0.9497 $\pm$ 0.0558 & 0.7246 $\pm$ 0.2703 & 0.2822 $\pm$ 0.1831 & \textbf{0.2522 $\pm$ 0.1907} & 1.6796 $\pm$ 2.4661 \\
        BA           & \textbf{0.7483 $\pm$ 0.0941} & 0.7480 $\pm$ 0.0948 & 0.6825 $\pm$ 0.2143 & \textbf{0.9461 $\pm$ 0.0478} & 0.9338 $\pm$ 0.0647 & 0.8704 $\pm$ 0.2601 & \textbf{0.4522 $\pm$ 0.2779} & 0.4895 $\pm$ 0.3258 & 0.5992 $\pm$ 0.8116 \\
        PCA (R)      & 0.8329 $\pm$ 0.1078 & \textbf{0.8333 $\pm$ 0.1330} & 0.8106 $\pm$ 0.1375 & 0.8849 $\pm$ 0.1160 & 0.8510 $\pm$ 0.1440 & \textbf{0.8677 $\pm$ 0.1580} & \textbf{0.5269 $\pm$ 0.8478} & 0.5880 $\pm$ 0.7596 & 0.7176 $\pm$ 1.1256 \\
        PCA (L)      & \textbf{0.8803 $\pm$ 0.0714} & 0.8763 $\pm$ 0.0857 & 0.8136 $\pm$ 0.1303 & \textbf{0.9334 $\pm$ 0.0763} & 0.9003 $\pm$ 0.1087 & 0.8800 $\pm$ 0.1204 & \textbf{0.2427 $\pm$ 0.4130} & 0.2883 $\pm$ 0.3569 & 0.5235 $\pm$ 0.6201 \\
        PComm (R)    & 0.5961 $\pm$ 0.3394 & \textbf{0.7269 $\pm$ 0.3181} & 0.2400 $\pm$ 0.3482 & 0.6453 $\pm$ 0.3853 & \textbf{0.7894 $\pm$ 0.3390} & 0.2998 $\pm$ 0.4174 & 0.7700 $\pm$ 1.0989 & \textbf{0.1989 $\pm$ 0.2544} & 0.2932 $\pm$ 0.2373 \\
        PComm (L)    & \textbf{0.6640 $\pm$ 0.2967} & 0.5773 $\pm$ 0.4203 & 0.2581 $\pm$ 0.3659 & \textbf{0.7301 $\pm$ 0.3148} & 0.6164 $\pm$ 0.4435 & 0.3137 $\pm$ 0.4463 & 0.4008 $\pm$ 0.4487 & \textbf{0.1290 $\pm$ 0.1379} & 0.2382 $\pm$ 0.0700 \\
        Other artery & \textbf{0.6037 $\pm$ 0.1083} & 0.5963 $\pm$ 0.1123 & --- & \textbf{0.6613 $\pm$ 0.0982} & 0.6452 $\pm$ 0.0998 & --- & \textbf{2.2988 $\pm$ 2.1245} & 2.3860 $\pm$ 2.0281 & --- \\
        Vein         & \textbf{0.9216 $\pm$ 0.0318} & --- & --- & \textbf{0.9738 $\pm$ 0.0254} & --- & --- & \textbf{0.1057 $\pm$ 0.0482} & --- & --- \\
        \bottomrule
    \end{tabular}}
\end{table}

\begin{table}[H]
\scriptsize
\centering
\caption{Aggregate segmentation performance label configurations using Dice Loss/Cross-Entropy (CE) Loss, and Skeleton Recall Loss on the TopCoW dataset (mean $\pm$ std).}
\label{tab:topcow_aggregate_combined}
\begin{tabular}{lllp{0.1cm} ccc}
\toprule
\textbf{Loss} & \textbf{Configuration} & \textbf{Evaluated vessel} && \textbf{mDC $\uparrow$} & \textbf{tSens $\uparrow$} & \textbf{adHD $\downarrow$} \\
\midrule
\multirow{3}{*}{\textbf{Dice/CE}}
    & \textbf{All labels}
        & Artery && $0.9457 \pm 0.0359$ & $0.9837 \pm 0.0208$ & $0.0533 \pm 0.0401$ \\
\cmidrule(lr){2-7}
    & \textbf{All arteries}
        & Artery && $0.9581 \pm 0.0331$ & $0.9846 \pm 0.0220$ & $0.0427 \pm 0.0392$ \\
\cmidrule(lr){2-7}
    & \textbf{Names arteries}
        & Artery && $0.8865 \pm 0.0637$ & $0.9510 \pm 0.0457$ & $0.1480 \pm 0.1677$ \\
\midrule
\multirow{2}{*}{\textbf{Skeleton}}
    & \textbf{All labels}
        & Artery && $\textbf{0.9638} \pm 0.0273$ & $\textbf{0.9909} \pm 0.0153$ & $\textbf{0.0334} \pm 0.0288$ \\
\cmidrule(lr){2-7}
    & \textbf{All arteries}
        & Artery && $0.9496 \pm 0.0334$ & $0.9836 \pm 0.0218$ & $0.0508 \pm 0.0399$ \\
\bottomrule
\end{tabular}
\end{table}

\begin{table}[H]
    \centering
    \scriptsize
    \setlength{\tabcolsep}{4pt}
    \caption{Per-label segmentation results across label configurations trained with Dice Loss and Cross-Entropy (CE) Loss (mean $\pm$ std) on the TopCoW dataset. }
    \label{tab:topcow_per_label_dice}
    \resizebox{\textwidth}{!}{%
    \begin{tabular}{l ccc ccc ccc}
        \toprule
        & \multicolumn{3}{c}{\textbf{mDC $\uparrow$}} 
        & \multicolumn{3}{c}{\textbf{tSens $\uparrow$}} 
        & \multicolumn{3}{c}{\textbf{adHD $\downarrow$}} \\
        \cmidrule(lr){2-4} \cmidrule(lr){5-7} \cmidrule(lr){8-10}
        \textbf{Label} 
        & \textbf{All labels} & \textbf{All arteries} & \textbf{Named arteries}
        & \textbf{All labels} & \textbf{All arteries} & \textbf{Named arteries}
        & \textbf{All labels} & \textbf{All arteries} & \textbf{Named arteries} \\
        \midrule
        ICA (R)      & \textbf{0.7618 $\pm$ 0.0762} & 0.7449 $\pm$ 0.1509 & 0.7019 $\pm$ 0.0903 & \textbf{0.8904 $\pm$ 0.0874} & 0.8606 $\pm$ 0.1573 & 0.8703 $\pm$ 0.1062 & \textbf{0.4156 $\pm$ 0.1885} & 0.5784 $\pm$ 0.9936 & 0.5479 $\pm$ 0.3228 \\
        ICA (L)      & \textbf{0.7736 $\pm$ 0.0690} & 0.7659 $\pm$ 0.1504 & 0.7003 $\pm$ 0.0995 & \textbf{0.8910 $\pm$ 0.0954} & 0.8676 $\pm$ 0.1784 & 0.8681 $\pm$ 0.1202 & \textbf{0.3696 $\pm$ 0.1592} & 0.5735 $\pm$ 1.5193 & 0.4988 $\pm$ 0.2635 \\
        M1 (R)       & 0.8038 $\pm$ 0.1589 & \textbf{0.8345 $\pm$ 0.1550} & 0.7739 $\pm$ 0.1730 & 0.8143 $\pm$ 0.2082 & \textbf{0.8343 $\pm$ 0.1988} & 0.8143 $\pm$ 0.2189 & 0.6254 $\pm$ 0.9330 & \textbf{0.5308 $\pm$ 0.8611} & 0.7104 $\pm$ 1.1389 \\
        M1 (L)       & 0.8144 $\pm$ 0.1706 & \textbf{0.8251 $\pm$ 0.1718} & 0.8189 $\pm$ 0.1530 & 0.8207 $\pm$ 0.2163 & 0.8362 $\pm$ 0.2155 & \textbf{0.8585 $\pm$ 0.1992} & 0.6168 $\pm$ 0.8452 & 0.5641 $\pm$ 0.8045 & \textbf{0.4992 $\pm$ 0.8107} \\
        ACA (R)      & \textbf{0.9084 $\pm$ 0.1059} & 0.9014 $\pm$ 0.1168 & 0.8885 $\pm$ 0.1029 & 0.9524 $\pm$ 0.0963 & 0.9348 $\pm$ 0.1085 & \textbf{0.9532 $\pm$ 0.1035} & \textbf{0.1016 $\pm$ 0.1825} & 0.1343 $\pm$ 0.2467 & 0.1524 $\pm$ 0.3281 \\
        ACA (L)      & 0.8539 $\pm$ 0.1431 & \textbf{0.8718 $\pm$ 0.1342} & 0.7934 $\pm$ 0.1670 & 0.9068 $\pm$ 0.1406 & \textbf{0.9159 $\pm$ 0.1285} & 0.8769 $\pm$ 0.1705 & 0.1915 $\pm$ 0.2687 & \textbf{0.1691 $\pm$ 0.2683} & 0.3654 $\pm$ 0.9038 \\
        AComm        & 0.4951 $\pm$ 0.3619 & \textbf{0.5346 $\pm$ 0.3481} & 0.2729 $\pm$ 0.3298 & 0.5365 $\pm$ 0.4651 & \textbf{0.5755 $\pm$ 0.4547} & 0.3350 $\pm$ 0.4421 & \textbf{0.3485 $\pm$ 0.5930} & 0.3748 $\pm$ 0.5820 & 0.4066 $\pm$ 0.3597 \\
        BA           & \textbf{0.7027 $\pm$ 0.1116} & 0.7004 $\pm$ 0.0985 & 0.5834 $\pm$ 0.1715 & \textbf{0.8833 $\pm$ 0.1371} & 0.8806 $\pm$ 0.1384 & 0.7965 $\pm$ 0.2459 & \textbf{0.5042 $\pm$ 0.3781} & 0.5103 $\pm$ 0.3424 & 0.8704 $\pm$ 1.5262 \\
        PCA (R)      & 0.9274 $\pm$ 0.0859 & \textbf{0.9460 $\pm$ 0.0848} & 0.9064 $\pm$ 0.1024 & 0.9720 $\pm$ 0.0852 & \textbf{0.9774 $\pm$ 0.0807} & 0.9641 $\pm$ 0.1066 & 0.0729 $\pm$ 0.1738 & \textbf{0.0576 $\pm$ 0.1721} & 0.1248 $\pm$ 0.3163 \\
        PCA (L)      & 0.9359 $\pm$ 0.0623 & \textbf{0.9499 $\pm$ 0.0521} & 0.9025 $\pm$ 0.1082 & \textbf{0.9834 $\pm$ 0.0486} & \textbf{0.9834 $\pm$ 0.0431} & 0.9603 $\pm$ 0.1099 & 0.0580 $\pm$ 0.0912 & \textbf{0.0451 $\pm$ 0.0749} & 0.1489 $\pm$ 0.4548 \\
        PComm (R)    & 0.6771 $\pm$ 0.3220 & \textbf{0.7432 $\pm$ 0.2812} & 0.3084 $\pm$ 0.3515 & 0.7321 $\pm$ 0.3544 & \textbf{0.8093 $\pm$ 0.2960} & 0.3529 $\pm$ 0.4012 & 0.3435 $\pm$ 0.6273 & \textbf{0.3198 $\pm$ 0.6174} & 0.7525 $\pm$ 0.8806 \\
        PComm (L)    & \textbf{0.6632 $\pm$ 0.2779} & 0.6564 $\pm$ 0.3361 & 0.3491 $\pm$ 0.3409 & \textbf{0.7679 $\pm$ 0.2991} & 0.7373 $\pm$ 0.3626 & 0.4075 $\pm$ 0.3990 & \textbf{0.3936 $\pm$ 0.5131} & 0.4670 $\pm$ 0.9512 & 0.9355 $\pm$ 1.1715 \\
        Other artery & 0.0000 $\pm$ 0.0000 & \textbf{0.0180 $\pm$ 0.0675} & --- & 0.0000 $\pm$ 0.0000 & \textbf{0.0200 $\pm$ 0.0748} & --- & 16.9444 $\pm$ 4.0454 & \textbf{15.4374 $\pm$ 5.3517} & --- \\
        \bottomrule
    \end{tabular}}
\end{table}

\begin{table}[H]
    \centering
    \scriptsize
    \setlength{\tabcolsep}{4pt}
    \caption{Per-label segmentation results for the model trained with Skeleton Recall Loss on the TopCoW  dataset (mean $\pm$ std).}
    \label{tab:topcow_per_label_skeleton}
    \resizebox{\textwidth}{!}{%
    \begin{tabular}{l cc cc cc}
        \toprule
        & \multicolumn{2}{c}{\textbf{mDC $\uparrow$}} 
        & \multicolumn{2}{c}{\textbf{tSens $\uparrow$}} 
        & \multicolumn{2}{c}{\textbf{adHD $\downarrow$}} \\
        \cmidrule(lr){2-3} \cmidrule(lr){4-5} \cmidrule(lr){6-7}
        \textbf{Label} 
        & \textbf{All labels} & \textbf{All arteries}
        & \textbf{All labels} & \textbf{All arteries}
        & \textbf{All labels} & \textbf{All arteries} \\
        \midrule
        ICA (R)      & \textbf{0.7634 $\pm$ 0.0711} & $0.7488 \pm 0.0721$ & $0.8646 \pm 0.0847$ & \textbf{0.8662 $\pm$ 0.0858} & \textbf{0.4279 $\pm$ 0.1922} & $0.4596 \pm 0.1893$ \\
        ICA (L)      & \textbf{0.7770 $\pm$ 0.0648} & $0.7600 \pm 0.0716$ & \textbf{0.8689 $\pm$ 0.0939} & $0.8624 \pm 0.1016$ & \textbf{0.3807 $\pm$ 0.1606} & $0.4095 \pm 0.1746$ \\
        M1 (R)       & $0.8238 \pm 0.1600$ & \textbf{0.8292 $\pm$ 0.1560} & $0.8258 \pm 0.2048$ & \textbf{0.8384 $\pm$ 0.1998} & $0.6030 \pm 0.9330$ & \textbf{0.5573 $\pm$ 0.8716} \\
        M1 (L)       & $0.8442 \pm 0.1561$ & \textbf{0.8497 $\pm$ 0.1557} & $0.8558 \pm 0.1955$ & \textbf{0.8649 $\pm$ 0.1951} & $0.4811 \pm 0.6766$ & \textbf{0.4669 $\pm$ 0.7006} \\
        ACA (R)      & \textbf{0.8965 $\pm$ 0.1155} & $0.8867 \pm 0.1211$ & \textbf{0.9354 $\pm$ 0.1148} & $0.9314 \pm 0.1184$ & \textbf{0.1191 $\pm$ 0.1967} & $0.1401 \pm 0.2409$ \\
        ACA (L)      & $0.8536 \pm 0.1273$ & \textbf{0.8615 $\pm$ 0.1222} & $0.8951 \pm 0.1265$ & \textbf{0.9029 $\pm$ 0.1186} & $0.1729 \pm 0.2370$ & \textbf{0.1575 $\pm$ 0.2143} \\
        AComm        & \textbf{0.6460 $\pm$ 0.3190} & $0.6044 \pm 0.3317$ & \textbf{0.6890 $\pm$ 0.4303} & $0.6488 \pm 0.4463$ & \textbf{0.2728 $\pm$ 0.4477} & $0.2860 \pm 0.4698$ \\
        BA           & \textbf{0.7039 $\pm$ 0.1041} & $0.6854 \pm 0.1059$ & \textbf{0.8696 $\pm$ 0.1412} & $0.8631 \pm 0.1369$ & \textbf{0.4928 $\pm$ 0.2589} & $0.5285 \pm 0.2666$ \\
        PCA (R)      & \textbf{0.9411 $\pm$ 0.0801} & $0.9288 \pm 0.0937$ & \textbf{0.9733 $\pm$ 0.0822} & $0.9696 \pm 0.0956$ & \textbf{0.0613 $\pm$ 0.1462} & $0.0723 \pm 0.1722$ \\
        PCA (L)      & \textbf{0.9489 $\pm$ 0.0626} & $0.9478 \pm 0.0544$ & $0.9776 \pm 0.0550$ & \textbf{0.9842 $\pm$ 0.0458} & $0.0509 \pm 0.1027$ & \textbf{0.0496 $\pm$ 0.0966} \\
        PComm (R)    & \textbf{0.6623 $\pm$ 0.3184} & $0.6180 \pm 0.3382$ & \textbf{0.7404 $\pm$ 0.3450} & $0.6991 \pm 0.3740$ & \textbf{0.2589 $\pm$ 0.3190} & $0.3990 \pm 0.6599$ \\
        PComm (L)    & \textbf{0.6776 $\pm$ 0.2994} & $0.6670 \pm 0.2957$ & $0.7607 \pm 0.3249$ & \textbf{0.7608 $\pm$ 0.3206} & \textbf{0.4115 $\pm$ 0.6359} & $0.4382 \pm 0.6836$ \\
        Other artery & \textbf{0.0000 $\pm$ 0.0000} & \textbf{0.0000 $\pm$ 0.0000} & \textbf{0.0000 $\pm$ 0.0000} & \textbf{0.0000 $\pm$ 0.0000} & $14.3025 \pm 4.7048$ & \textbf{14.0284 $\pm$ 3.1399} \\
        \bottomrule
    \end{tabular}}
\end{table}

\begin{table}[H]
\centering
\caption{Aggregate segmentation performance of multiphase and arterial phase only models trained with the Skeleton Recall Loss on the TopBrain and TopCoW datasets (mean $\pm$ std).}
\label{tab:arterial_phase}
\resizebox{\textwidth}{!}{%
\begin{tabular}{lllp{0.1cm} ccc}
\toprule
\textbf{Dataset} & \textbf{Configuration} & \textbf{Evaluated vessel} && \textbf{mDC $\uparrow$} & \textbf{tSens $\uparrow$} & \textbf{adHD $\downarrow$} \\
\midrule
\multirow{6}{*}{\textbf{TopBrain}}
    & \multirow{3}{*}{\textbf{All labels}}
        & All vessel && $\mathbf{0.9381 \pm 0.0217}$ & $\mathbf{0.9740 \pm 0.0120}$ & $\mathbf{0.1008 \pm 0.0511}$ \\
    & & Artery      && $\mathbf{0.9025 \pm 0.0373}$ & $\mathbf{0.9477 \pm 0.0228}$ & $\mathbf{0.1945 \pm 0.1176}$ \\
    & & Vein        && $0.9422 \pm 0.0334$ & $\mathbf{0.9724 \pm 0.0253}$ & $0.0758 \pm 0.0435$ \\
\cmidrule(lr){2-7}
    & \multirow{3}{*}{\textbf{All labels (arterial phase only)}}
        & All vessel && $0.9292 \pm 0.0232$ & $0.9532 \pm 0.0119$ & $0.1261 \pm 0.0422$ \\
    & & Artery      && $0.8616 \pm 0.0489$ & $0.9310 \pm 0.0229$ & $0.3380 \pm 0.4021$ \\
    & & Vein        && $\mathbf{0.9727 \pm 0.0338}$ & $0.9527 \pm 0.0497$ & $\mathbf{0.0568 \pm 0.0624}$ \\
\midrule
\multirow{2}{*}{\textbf{TopCoW}}
    & \textbf{All labels}
        & Artery     && $\mathbf{0.9638 \pm 0.0273}$ & $\mathbf{0.9909 \pm 0.0153}$ & $\mathbf{0.0334 \pm 0.0288}$ \\
\cmidrule(lr){2-7}
    & \textbf{All labels (arterial phase only)}
        & Artery     && $0.9472 \pm 0.0415$ & $0.9760 \pm 0.0365$ & $0.0710 \pm 0.1431$ \\
\bottomrule
\end{tabular}%
}
\end{table}

\begin{table}[H]
\centering
\caption{Aggregate segmentation performance of three models evaluated on 5 held-out TopBrain cases (mean $\pm$ std). All models use Skeleton Recall Loss.}
\label{tab:cross_dataset_topbrain}
\resizebox{\textwidth}{!}{%
\begin{tabular}{llp{0.1cm} ccc}
\toprule
\textbf{Training set} & \textbf{Evaluated vessel} && \textbf{mDC $\uparrow$} & \textbf{tSens $\uparrow$} & \textbf{adHD $\downarrow$} \\
\midrule
\multirow{3}{*}{\textbf{\shortstack[l]{SemanticVessel \\ (360 scans, multiphase)}}}
    & All vessel && $\mathbf{0.9545 \pm 0.0049}$ & $\mathbf{0.9792 \pm 0.0031}$ & $\mathbf{0.0754 \pm 0.0186}$ \\
    & Artery     && $\mathbf{0.8966 \pm 0.0553}$ & $\mathbf{0.9532 \pm 0.0121}$ & $\mathbf{0.2036 \pm 0.1431}$ \\
    & Vein       && $0.9659 \pm 0.0039$ & $\mathbf{0.9740 \pm 0.0270}$ & $0.0489 \pm 0.0144$ \\
\midrule
\multirow{3}{*}{\textbf{\shortstack[l]{SemanticVessel \\ (41 scans, arterial \\ phase only)}}}
    & All vessel && $0.9402 \pm 0.0185$ & $0.9485 \pm 0.0129$ & $0.1091 \pm 0.0406$ \\
    & Artery     && $0.8353 \pm 0.0809$ & $0.9197 \pm 0.0289$ & $0.6363 \pm 0.8027$ \\
    & Vein       && $\mathbf{0.9895 \pm 0.0076}$ & $0.9516 \pm 0.0406$ & $\mathbf{0.0362 \pm 0.0384}$ \\
\midrule
\multirow{3}{*}{\textbf{TopBrain (20 scans)}}
    & All vessel && $0.8337 \pm 0.0429$ & $0.8292 \pm 0.0325$ & $0.3852 \pm 0.1179$ \\
    & Artery     && $0.7804 \pm 0.0449$ & $0.7980 \pm 0.0396$ & $0.5103 \pm 0.1596$ \\
    & Vein       && $0.8891 \pm 0.0672$ & $0.8734 \pm 0.0676$ & $0.3044 \pm 0.2753$ \\
\bottomrule
\end{tabular}%
}
\end{table}

\begin{table}[H]
\centering
\caption{Aggregate segmentation performance of three models evaluated on 5 held-out SemanticVessel arterial phase cases (mean $\pm$ std). All models use Skeleton Recall Loss.}
\label{tab:cross_dataset_semantic}
\resizebox{\textwidth}{!}{%
\begin{tabular}{llp{0.1cm} ccc}
\toprule
\textbf{Training set} & \textbf{Evaluated vessel} && \textbf{mDC $\uparrow$} & \textbf{tSens $\uparrow$} & \textbf{adHD $\downarrow$} \\
\midrule
\multirow{3}{*}{\textbf{\shortstack[l]{SemanticVessel \\ (360 scans, multiphase)}}}
    & All vessel && $\mathbf{0.7555 \pm 0.1336}$ & $\mathbf{0.8158 \pm 0.0784}$ & $\mathbf{0.4881 \pm 0.2890}$ \\
    & Artery     && $\mathbf{0.8599 \pm 0.0489}$ & $\mathbf{0.9166 \pm 0.0161}$ & $\mathbf{0.2398 \pm 0.0804}$ \\
    & Vein       && $\mathbf{0.7185 \pm 0.1676}$ & $\mathbf{0.7426 \pm 0.1235}$ & $\mathbf{0.6725 \pm 0.4369}$ \\
\midrule
\multirow{3}{*}{\textbf{\shortstack[l]{SemanticVessel \\ (41 scans, arterial \\ phase only)}}}
    & All vessel && $0.6550 \pm 0.1676$ & $0.6805 \pm 0.1302$ & $1.0330 \pm 0.5532$ \\
    & Artery     && $0.7427 \pm 0.1735$ & $0.7872 \pm 0.1347$ & $0.8806 \pm 0.6583$ \\
    & Vein       && $0.6023 \pm 0.1767$ & $0.5156 \pm 0.1250$ & $1.7257 \pm 0.9209$ \\
\midrule
\multirow{3}{*}{\textbf{TopBrain (20 scans)}}
    & All vessel && $0.1197 \pm 0.0440$ & $0.1555 \pm 0.0365$ & $12.8017 \pm 1.0111$ \\
    & Artery     && $0.2447 \pm 0.0595$ & $0.3111 \pm 0.0690$ & $10.2010 \pm 1.4143$ \\
    & Vein       && $0.0660 \pm 0.0305$ & $0.0660 \pm 0.0226$ & $22.7483 \pm 2.5208$ \\
\bottomrule
\end{tabular}%
}
\end{table}

\begin{table}[H]
\centering
\caption{Per-label segmentation results of three models evaluated on 5 held-out SemanticVessel arterial phase cases (mean $\pm$ std). All models use Skeleton Recall Loss. Multi = SemanticVessel multiphase (360 scans); Single = SemanticVessel arterial phase only (41 scans); TopBrain = TopBrain trained (20 scans).}
\label{tab:per_label_semantic_vessel}
\resizebox{\textwidth}{!}{%
\begin{tabular}{l ccc ccc ccc}
\toprule
\multirow{2}{*}{\textbf{Label}} & \multicolumn{3}{c}{\textbf{mDC $\uparrow$}} & \multicolumn{3}{c}{\textbf{tSens $\uparrow$}} & \multicolumn{3}{c}{\textbf{adHD $\downarrow$}} \\
\cmidrule(lr){2-4} \cmidrule(lr){5-7} \cmidrule(lr){8-10}
& \textbf{Multi} & \textbf{Single} & \textbf{TopBrain} & \textbf{Multi} & \textbf{Single} & \textbf{TopBrain} & \textbf{Multi} & \textbf{Single} & \textbf{TopBrain} \\
\midrule
ICA (R)      & $\mathbf{0.9611 \pm 0.0294}$ & $0.8905 \pm 0.1267$ & $0.0806 \pm 0.0203$ & $\mathbf{1.0000 \pm 0.0000}$ & $0.9842 \pm 0.0265$ & $0.1157 \pm 0.0254$ & $\mathbf{0.0715 \pm 0.0514}$ & $0.2020 \pm 0.2374$ & $23.8944 \pm 1.5023$ \\
ICA (L)      & $\mathbf{0.9646 \pm 0.0275}$ & $0.8856 \pm 0.1579$ & $0.0930 \pm 0.0193$ & $\mathbf{1.0000 \pm 0.0000}$ & $0.9565 \pm 0.0870$ & $0.1404 \pm 0.0255$ & $\mathbf{0.0623 \pm 0.0488}$ & $0.2360 \pm 0.3508$ & $22.8023 \pm 1.2483$ \\
M1 (R)       & $\mathbf{0.9004 \pm 0.0343}$ & $0.8546 \pm 0.0805$ & $0.5299 \pm 0.1626$ & $\mathbf{0.9677 \pm 0.0408}$ & $0.8956 \pm 0.1127$ & $0.7917 \pm 0.1642$ & $\mathbf{0.1442 \pm 0.0898}$ & $0.2352 \pm 0.1304$ & $0.9265 \pm 0.5380$ \\
M1 (L)       & $0.8763 \pm 0.1171$ & $\mathbf{0.9205 \pm 0.0804}$ & $0.7054 \pm 0.1326$ & $0.9789 \pm 0.0421$ & $\mathbf{1.0000 \pm 0.0000}$ & $0.9339 \pm 0.0770$ & $0.2166 \pm 0.2898$ & $\mathbf{0.1246 \pm 0.1387}$ & $0.4916 \pm 0.3388$ \\
M2+ (R)      & $\mathbf{0.8886 \pm 0.0927}$ & $0.7516 \pm 0.2398$ & $0.2209 \pm 0.0960$ & $\mathbf{0.9733 \pm 0.0298}$ & $0.8120 \pm 0.1756$ & $0.3170 \pm 0.0976$ & $\mathbf{0.1294 \pm 0.1167}$ & $1.0920 \pm 1.2461$ & $7.7854 \pm 2.1956$ \\
M2+ (L)      & $\mathbf{0.8966 \pm 0.0832}$ & $0.7328 \pm 0.2221$ & $0.3166 \pm 0.1200$ & $\mathbf{0.9807 \pm 0.0197}$ & $0.7899 \pm 0.1795$ & $0.4210 \pm 0.1353$ & $\mathbf{0.1064 \pm 0.0922}$ & $1.1561 \pm 1.3066$ & $6.6268 \pm 2.6073$ \\
ACA (R)      & $\mathbf{0.8939 \pm 0.0478}$ & $0.7330 \pm 0.1885$ & $0.3645 \pm 0.1131$ & $\mathbf{0.9516 \pm 0.0385}$ & $0.7840 \pm 0.1610$ & $0.3988 \pm 0.1650$ & $\mathbf{0.1273 \pm 0.0667}$ & $0.7421 \pm 0.8062$ & $10.9800 \pm 6.0148$ \\
ACA (L)      & $\mathbf{0.8117 \pm 0.0239}$ & $0.7001 \pm 0.1547$ & $0.2541 \pm 0.1537$ & $\mathbf{0.8880 \pm 0.0504}$ & $0.7770 \pm 0.1151$ & $0.2659 \pm 0.1640$ & $\mathbf{0.2782 \pm 0.0936}$ & $0.6593 \pm 0.5028$ & $12.2836 \pm 6.6502$ \\
AComm        & $0.5266 \pm 0.4390$ & $\mathbf{0.5620 \pm 0.3562}$ & $0.0000 \pm 0.0000$ & $0.5714 \pm 0.4695$ & $\mathbf{0.6737 \pm 0.4163}$ & $0.0000 \pm 0.0000$ & $\mathbf{0.0687 \pm 0.0642}$ & $0.3258 \pm 0.2847$ & $-$ \\
VA (R)       & $\mathbf{0.9385 \pm 0.0379}$ & $0.6345 \pm 0.3716$ & $0.4997 \pm 0.3024$ & $\mathbf{0.9875 \pm 0.0125}$ & $0.7321 \pm 0.4237$ & $0.6209 \pm 0.3736$ & $\mathbf{0.0720 \pm 0.0465}$ & $0.2253 \pm 0.1405$ & $0.9674 \pm 0.6131$ \\
VA (L)       & $\mathbf{0.9327 \pm 0.0431}$ & $0.6933 \pm 0.2728$ & $0.7704 \pm 0.1274$ & $\mathbf{0.9649 \pm 0.0608}$ & $0.8184 \pm 0.1928$ & $0.8781 \pm 0.0956$ & $\mathbf{0.0975 \pm 0.0741}$ & $0.7897 \pm 0.9585$ & $0.5998 \pm 0.4842$ \\
BA           & $\mathbf{0.6888 \pm 0.3458}$ & $0.5951 \pm 0.3800$ & $0.5915 \pm 0.4750$ & $\mathbf{0.7562 \pm 0.3804}$ & $0.7226 \pm 0.3839$ & $0.5939 \pm 0.4795$ & $\mathbf{3.9406 \pm 7.3457}$ & $4.5054 \pm 7.3539$ & $4.4118 \pm 7.5983$ \\
PCA (R)      & $\mathbf{0.8397 \pm 0.1303}$ & $0.7572 \pm 0.2432$ & $0.3783 \pm 0.1619$ & $\mathbf{0.9428 \pm 0.0643}$ & $0.8475 \pm 0.1915$ & $0.4502 \pm 0.1476$ & $\mathbf{2.4845 \pm 4.7551}$ & $3.1757 \pm 4.5586$ & $6.8951 \pm 4.4078$ \\
PCA (L)      & $\mathbf{0.8457 \pm 0.1063}$ & $0.7617 \pm 0.1699$ & $0.4458 \pm 0.1642$ & $\mathbf{0.9537 \pm 0.0662}$ & $0.8463 \pm 0.1067$ & $0.5636 \pm 0.0894$ & $\mathbf{0.8524 \pm 1.4972}$ & $1.1147 \pm 1.3122$ & $3.2699 \pm 1.3291$ \\
PComm (R)    & $\mathbf{0.8529 \pm 0.1524}$ & $0.7019 \pm 0.1988$ & $0.0000 \pm 0.0000$ & $\mathbf{0.8463 \pm 0.1955}$ & $0.7581 \pm 0.1811$ & $0.0000 \pm 0.0000$ & $\mathbf{0.2131 \pm 0.2726}$ & $0.3092 \pm 0.2201$ & $-$ \\
PComm (L)    & $\mathbf{0.9247 \pm 0.0277}$ & $0.6797 \pm 0.2042$ & $0.0000 \pm 0.0000$ & $\mathbf{1.0000 \pm 0.0000}$ & $0.9231 \pm 0.0769$ & $0.0000 \pm 0.0000$ & $\mathbf{0.0735 \pm 0.0307}$ & $0.3169 \pm 0.1855$ & $-$ \\
Other artery & $\mathbf{0.6713 \pm 0.1411}$ & $0.4242 \pm 0.2263$ & $0.0753 \pm 0.0438$ & $\mathbf{0.7694 \pm 0.0503}$ & $0.4634 \pm 0.1963$ & $0.0999 \pm 0.0464$ & $\mathbf{0.6596 \pm 0.2351}$ & $2.8469 \pm 1.4715$ & $28.7639 \pm 3.6114$ \\
Vein         & $\mathbf{0.7185 \pm 0.1676}$ & $0.6023 \pm 0.1767$ & $0.0660 \pm 0.0305$ & $\mathbf{0.7426 \pm 0.1235}$ & $0.5156 \pm 0.1250$ & $0.0660 \pm 0.0226$ & $\mathbf{0.6725 \pm 0.4369}$ & $1.7257 \pm 0.9209$ & $22.7483 \pm 2.5208$ \\
\bottomrule
\end{tabular}%
}
\end{table}

\begin{table}[H]
\centering
\caption{Per-label segmentation results of three models evaluated on 5 held-out TopBrain cases (mean $\pm$ std). All models use Skeleton Recall Loss. Multi = SemanticVessel multiphase (360 scans); Single = SemanticVessel arterial phase only (41 scans); TopBrain = TopBrain trained (20 scans).}
\label{tab:per_label_topbrain}
\resizebox{\textwidth}{!}{%
\begin{tabular}{l ccc ccc ccc}
\toprule
\multirow{2}{*}{\textbf{Label}} & \multicolumn{3}{c}{\textbf{mDC $\uparrow$}} & \multicolumn{3}{c}{\textbf{tSens $\uparrow$}} & \multicolumn{3}{c}{\textbf{adHD $\downarrow$}} \\
\cmidrule(lr){2-4} \cmidrule(lr){5-7} \cmidrule(lr){8-10}
& \textbf{Multi} & \textbf{Single} & \textbf{TopBrain} & \textbf{Multi} & \textbf{Single} & \textbf{TopBrain} & \textbf{Multi} & \textbf{Single} & \textbf{TopBrain} \\
\midrule
ICA (R)      & $0.7670 \pm 0.1069$ & $0.6717 \pm 0.0369$ & $\mathbf{0.8355 \pm 0.0813}$ & $0.9165 \pm 0.0526$ & $0.8645 \pm 0.0556$ & $\mathbf{0.9778 \pm 0.0444}$ & $0.3959 \pm 0.2262$ & $0.5543 \pm 0.1691$ & $\mathbf{0.1989 \pm 0.1194}$ \\
ICA (L)      & $0.7543 \pm 0.0557$ & $0.7232 \pm 0.0610$ & $\mathbf{0.7981 \pm 0.0630}$ & $0.8563 \pm 0.1190$ & $0.8245 \pm 0.1143$ & $\mathbf{0.9684 \pm 0.0632}$ & $0.4499 \pm 0.1643$ & $0.4902 \pm 0.1712$ & $\mathbf{0.2877 \pm 0.1504}$ \\
M1 (R)       & $0.6354 \pm 0.1527$ & $0.6787 \pm 0.1053$ & $\mathbf{0.8455 \pm 0.0734}$ & $0.5652 \pm 0.2299$ & $0.6694 \pm 0.1842$ & $\mathbf{0.9459 \pm 0.0663}$ & $1.2723 \pm 0.7771$ & $0.9311 \pm 0.4969$ & $\mathbf{0.2591 \pm 0.2354}$ \\
M1 (L)       & $0.6351 \pm 0.0883$ & $\mathbf{0.7667 \pm 0.0890}$ & $0.7979 \pm 0.0644$ & $0.4912 \pm 0.1516$ & $0.5856 \pm 0.1214$ & $\mathbf{0.7426 \pm 0.1760}$ & $1.5722 \pm 0.7867$ & $\mathbf{0.7914 \pm 0.3078}$ & $0.4586 \pm 0.3618$ \\
M2+ (R)      & $\mathbf{0.9769 \pm 0.0115}$ & $0.9244 \pm 0.0449$ & $0.6817 \pm 0.1071$ & $\mathbf{0.9853 \pm 0.0100}$ & $0.8990 \pm 0.0522$ & $0.7043 \pm 0.1121$ & $\mathbf{0.0192 \pm 0.0126}$ & $0.2005 \pm 0.1091$ & $0.7536 \pm 0.2661$ \\
M2+ (L)      & $\mathbf{0.9830 \pm 0.0055}$ & $0.9037 \pm 0.0318$ & $0.6793 \pm 0.0483$ & $\mathbf{0.9976 \pm 0.0033}$ & $0.8943 \pm 0.0323$ & $0.7463 \pm 0.0612$ & $\mathbf{0.0106 \pm 0.0043}$ & $0.3572 \pm 0.1561$ & $0.9618 \pm 0.4028$ \\
ACA (R)      & $\mathbf{0.9025 \pm 0.0553}$ & $0.8395 \pm 0.1477$ & $0.8132 \pm 0.0641$ & $\mathbf{0.9735 \pm 0.0403}$ & $0.8858 \pm 0.1378$ & $0.8946 \pm 0.0911$ & $\mathbf{0.0774 \pm 0.0469}$ & $0.1994 \pm 0.2148$ & $0.3378 \pm 0.2517$ \\
ACA (L)      & $0.8006 \pm 0.1046$ & $\mathbf{0.8915 \pm 0.0572}$ & $0.7618 \pm 0.0953$ & $0.8499 \pm 0.1824$ & $\mathbf{0.9229 \pm 0.0520}$ & $0.8404 \pm 0.1356$ & $0.4261 \pm 0.5664$ & $\mathbf{0.1015 \pm 0.0613}$ & $0.3511 \pm 0.2236$ \\
AComm        & $\mathbf{0.7354 \pm 0.0830}$ & $0.4755 \pm 0.4756$ & $0.0000 \pm 0.0000$ & $\mathbf{0.7500 \pm 0.4330}$ & $0.5000 \pm 0.5000$ & $0.0000 \pm 0.0000$ & $\mathbf{0.2638 \pm 0.1137}$ & $0.0319 \pm 0.0085$ & $-$ \\
VA (R)       & $0.7561 \pm 0.1817$ & $0.5610 \pm 0.2515$ & $\mathbf{0.7600 \pm 0.0755}$ & $\mathbf{0.9197 \pm 0.1125}$ & $0.7273 \pm 0.2577$ & $0.8886 \pm 0.0662$ & $\mathbf{0.5446 \pm 0.6066}$ & $3.5044 \pm 5.0344$ & $0.3568 \pm 0.1153$ \\
VA (L)       & $\mathbf{0.8566 \pm 0.1024}$ & $0.6506 \pm 0.2604$ & $0.7349 \pm 0.2192$ & $\mathbf{0.9832 \pm 0.0283}$ & $0.7659 \pm 0.3125$ & $0.8711 \pm 0.2106$ & $\mathbf{0.1519 \pm 0.1175}$ & $3.6797 \pm 6.4809$ & $0.4349 \pm 0.4903$ \\
BA           & $0.7959 \pm 0.0852$ & $0.8172 \pm 0.0470$ & $\mathbf{0.8907 \pm 0.0448}$ & $0.9545 \pm 0.0594$ & $0.9683 \pm 0.0308$ & $\mathbf{1.0000 \pm 0.0000}$ & $0.3014 \pm 0.1633$ & $\mathbf{0.2673 \pm 0.0876}$ & $0.1344 \pm 0.0625$ \\
PCA (R)       & $0.9122 \pm 0.0371$ & $\mathbf{0.9354 \pm 0.0209}$ & $0.7515 \pm 0.1693$ & $0.9282 \pm 0.0426$ & $\mathbf{0.9425 \pm 0.0444}$ & $0.8399 \pm 0.1326$ & $0.1288 \pm 0.0911$ & $\mathbf{0.1186 \pm 0.0798}$ & $0.4676 \pm 0.3581$ \\
PCA (L)       & $0.8952 \pm 0.0407$ & $\mathbf{0.9350 \pm 0.0343}$ & $0.7323 \pm 0.0511$ & $0.9118 \pm 0.0480$ & $\mathbf{0.9517 \pm 0.0659}$ & $0.8206 \pm 0.1028$ & $0.2070 \pm 0.1148$ & $\mathbf{0.1159 \pm 0.1282}$ & $0.5586 \pm 0.4773$ \\
PComm (R)    & $\mathbf{0.4914 \pm 0.4914}$ & $0.6372 \pm 0.3456$ & $0.0000 \pm 0.0000$ & $\mathbf{0.5000 \pm 0.5000}$ & $0.6522 \pm 0.3478$ & $0.0000 \pm 0.0000$ & $\mathbf{0.0113 \pm 0.0000}$ & $0.5444 \pm 0.5319$ & $-$ \\
PComm (L)    & $\mathbf{0.8239 \pm 0.0000}$ & $0.0000 \pm 0.0000$ & $0.0000 \pm 0.0000$ & $\mathbf{0.9500 \pm 0.0000}$ & $0.0000 \pm 0.0000$ & $0.0000 \pm 0.0000$ & $\mathbf{0.1125 \pm 0.0000}$ & $-$ & $-$ \\
Other artery & $0.6452 \pm 0.2243$ & $\mathbf{0.7257 \pm 0.0588}$ & $0.6139 \pm 0.1976$ & $0.6541 \pm 0.1802$ & $\mathbf{0.7283 \pm 0.0459}$ & $0.6729 \pm 0.1698$ & $2.1274 \pm 2.5119$ & $\mathbf{0.9189 \pm 0.2169}$ & $1.1479 \pm 0.6128$ \\
Vein         & $0.9659 \pm 0.0039$ & $\mathbf{0.9895 \pm 0.0076}$ & $0.8891 \pm 0.0672$ & $\mathbf{0.9740 \pm 0.0270}$ & $0.9516 \pm 0.0406$ & $0.8734 \pm 0.0676$ & $0.0489 \pm 0.0144$ & $\mathbf{0.0362 \pm 0.0384}$ & $0.3044 \pm 0.2753$ \\
\bottomrule
\end{tabular}%
}
\end{table}
\end{document}